\documentclass{egpubl}
\usepackage{pg2020}
 
\SpecialIssuePaper         

\usepackage[T1]{fontenc}
\usepackage{dfadobe}  

\newcommand{\pk}[1]{{\color{dark_green}#1}}
\newcommand{\bin}[1]{{\color{red}#1}}
\newcommand{\uri}[1]{{\color{blue}#1}}

\newcommand{\binnew}[1]{{\color{pink}#1}}
\newcommand{\bincheck}[1]{{\color{brown}#1}}

\renewcommand{\pk}[1]{{\color{blue}#1}}
\renewcommand{\bin}[1]{{\color{blue}#1}}
\renewcommand{\uri}[1]{{\color{black}#1}}

\renewcommand{\binnew}[1]{{\color{blue}#1}}
\renewcommand{\bincheck}[1]{{\color{blue}#1}}

\renewcommand{\pk}[1]{{\color{black}#1}}
\renewcommand{\bin}[1]{{\color{black}#1}}

\renewcommand{\binnew}[1]{{\color{black}#1}}
\renewcommand{\bincheck}[1]{{\color{black}#1}}

\newcommand{\binTVCG}[1]{{\color{black}#1}}
\newcommand{\urin}[1]{{\color{black}#1}}
\newcommand{\binPG}[1]{{\color{black}#1}}
\newcommand{\rev}[1]{{\color{black}#1}}

\usepackage{bm}
\usepackage{amsmath, amsfonts, amssymb}
\usepackage{float}

\biberVersion
\BibtexOrBiblatex
\usepackage[backend=bibtex,bibstyle=EG,citestyle=alphabetic,backref=true]{biblatex} 
\electronicVersion
\PrintedOrElectronic

\ifpdf \usepackage[pdftex]{graphicx} \pdfcompresslevel=9
\else \usepackage[dvips]{graphicx} \fi

\usepackage{egweblnk} 

\title{Learning Elastic Constitutive Material and Damping Models}

\author[Bin Wang, Yuanmin Deng, Paul Kry, Uri Ascher, Hui Huang, Baoquan Chen]
{\parbox{\textwidth}{\centering 
Bin Wang$^{1}$,~
Yuanmin Deng\thanks{Joint first author}$^{2}$,~
Paul Kry$^{3}$,~
Uri Ascher$^{4}$,~
Hui Huang\thanks{Corresponding author (hhzhiyan@gmail.com)}$^{5}$~\orcid{0000-0003-3212-0544},~
Baoquan Chen\thanks{Corresponding author (baoquan@pku.edu.cn)}$^{6}$
}
        \\
{\parbox{\textwidth}{\centering 
$^1$Beijing Film Academy\quad
$^2$Shandong University\quad
$^3$McGill University\quad
$^4$University of British Columbia\quad
$^5$Shenzhen University\quad
$^6$Peking University  }
}
}

\begin{document}

\teaser{
\includegraphics[trim=3cm 1cm 3cm 1cm, width=.99\linewidth]{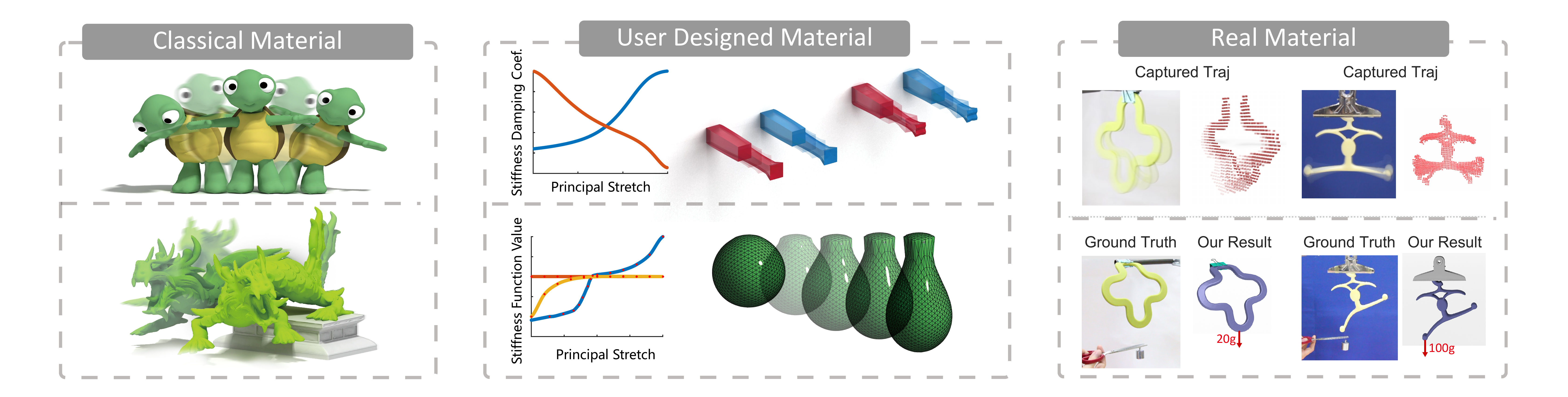}
\centering
\caption{	Our parametric material model learns a correction to a nominal material model from kinematic data alone, allowing us to accurately capture the nonlinearity of different constitutive material models. Left: classical nonlinear constitutive material. Middle: user designed elasticity and damping. Right: real world material. }
\label{fig:tease}
}

\maketitle
\begin{abstract}
   Commonly used linear and nonlinear constitutive material models in deformation simulation contain many simplifications and only cover a tiny part of possible material behavior. In this work we propose a framework for learning customized models of deformable materials from example surface trajectories. The key idea is to iteratively improve a correction to a nominal model of the elastic and damping properties of the object, which allows new forward simulations with the learned correction to more accurately predict the behavior of a given soft object. Space-time optimization is employed to identify gentle control forces with which we extract necessary data for model inference and to finally encapsulate the material correction into a compact parametric form. Furthermore, a patch based position constraint is proposed to tackle the challenge of handling incomplete and noisy observations arising in real-world examples. We demonstrate the effectiveness of our method with a set of synthetic examples, as well with data captured from real world homogeneous elastic objects. \\


\end{abstract}  
\section{Introduction}\label{sec:intro}

The simulation of deformable objects is ubiquitous in computer graphics and robotics research due to the large number of varied applications, including animation, movie making, medical treatment and manufacturing. These applications benefit from high fidelity deformation simulation, which is dependent upon the underlying constitutive material model.  Subsequently, the material parameters must be carefully tuned in order to fit empirical data. Data-driven methods have recently exhibited great potential in this direction. Advanced scanning technologies can be used to faithfully capture a deformation behavior under external force, and in turn the data can be used to estimate the parameters of the mathematical model.

However, there is currently no standard method for choosing the appropriate constitutive models, especially for large deformations, real-world materials, heterogeneous models, and artist designed cartoon physics.   The situation for modeling damping is even worse. Indeed, there is no agreement about models in the mechanical engineering literature which can describe various damping effects in a unified way.  Many use the Rayleigh model, which can be inadequate for visual purposes~\cite{Xu2017}. Based on these observations, in this paper we propose a more general material inference framework obtained by directly learning constitutive laws rather than fitting parameters from real-world motion data. Learning a constitutive model is challenging because: (1) the model should be able to encapsulate all the variations of material properties in a generic way; (2) there may be no obvious source of training data; and (3) real-world motion data is typically sparse and noisy.
We address these challenges and \rev{present}
the following contributions:
\vspace{-0.5em}
\begin{itemize}
\item 
a constitutive material model designed as a combination of empirical baseline model and a parametric correction; 
%
\item 
an inverse learning framework capable of learning a complex constitutive material from sparse motion trajectories; and
%
\item 
a differentiable patch based position constraint for probabilistic correspondences, which allows our system to work faithfully on real-world captured data.
\end{itemize}

Fig.~\ref{fig:tease} shows a preview of our approach and results.  We demonstrate the performance on several problems, including synthetic examples, coarsening applications, and captured data. The variety of results described in Section~\ref{sec:results} leads us to believe that our work \rev{contributes}
a useful technique and 
an important step in advancing data-driven 
constitutive elastodynamic force models.

\section{Related work}\label{sec:related}

Data-driven material parameter optimization offers great potential for computer graphics applications, such as fabrics, soft objects, and human organs and faces~\cite{Pai01,Schoner04,Becker07,Wang11,Miguel12,Bickel09}. Bickel et al.~\cite{Bickel09} fit material parameters with an incremental loading strategy to better approximate nonlinear strain-stress relationships. Wang et al.~\cite{Wang11} propose a piecewise linear elastic model to reproduce nonlinear, anisotropic stretching and bending of cloth.  Other appropaches directly optimize nonlinear stress-strain curves based on measurements~\cite{Miguel12}, and estimate internal friction~\cite{Miguel2013}.  Miguel et al.~\cite{Miguel2016} model example based cloth and elastic solid materials with energy functions.

A common limitation with previous methods is that they require a dense force displacement field. While Bhat et al.~\cite{Bhat03} avoid the need for force capture by using video tracking of cloth, they still assume a trivial cloth reference shape. Yang et al.~\cite{Yang2017} present a learning-based algorithm to recover material properties of cloth from videos, using training data sets generated by physics-based simulators.  Here, material type estimation is the focus due to inconsistencies between real and synthetic data and sparse material space sampling. Davis et al.~\cite{Davis2017} estimate material and damping properties by extracting small vibration modes from high-speed and regular frame-rate video.

Both Wang et al.~\cite{Wang2015} and the more recent Hahn et al.~\cite{Hahn19} estimate material properties from partially observed surface trajectories of an object's passive dynamics. A gradient-free Nelder-Mead optimization algorithm is employed by Wang et al., whereas Hahn et al.~compute the gradient with respect to material parameters using either direct sensitivity analysis or an adjoint state method.  Our work has a similar setting, but focuses on correcting the errors that arise when starting with simple elastic and damping force models
\uri{and differs in the methods used in the pipeline}.


Another popular trend in computer graphics is to directly fit parametric functions as a material description.  Xu et al.~\cite{Xu2015} provide a method to design isotropic and anisotropic (orthotropic) nonlinear solid elastic materials using a piecewise spline interface, which can provide local control on deformation behavior. A hyperelasticity model based on energy addends~\cite{Miguel2016} allows modeling of various nonlinear elasticity effects in a separable manner. Instead of explicitly modeling the stress-strain relationship, Martin et al.~\cite{Martin2011} and Schumacher et al.~\cite{Schumacher12} promote an art-directed approach to solid simulation, which constructs a manifold of preferred deformation examples, to which the object is guided.  

Also related are coarsening techniques, for instance, in applications for computational design for fabrication~\cite{Chen2015,Panetta2015,Chen2017,Chen:coarsen:2019}, where equivalent physics based models are important.  Kharevych et al.~\cite{Kharevych2009} take an energy based approach to coarsening composite elastic objects through the use of global harmonic displacements. Nesme et al.~\cite{Nesme2009} create nonlinear shape functions and projected fine-level mass, stiffness, and damping matrices to produce coarse composite elements, while Torres et al.~\cite{Torres2016} introduce an improved element based coarsening method that deals with corotation. 

In comparison to elasticity modeling, few publications have focused on the design of damping models. Xu et al.~\cite{Xu2017} present a method for designing anisotropic and/or nonlinear damping effects. Banderas et al.~\cite{Banderas18} model damping based on dissipation potentials using strain rate to control damping. (\uri{We} use a similar approach.) Targeting cloth hysteresis effects, Miguel et al.~\cite{Miguel2013} propose an internal friction model based on an augmented reparameterization of Dahl's model.  
In our work, inspired by Xu et al.~\cite{Xu2015}, we use principal stretches to formulate nonlinear elastic and damping force corrections.
However, instead of using 
splines, we adopt radial basis functions (RBF) to parameterize the material correction.



Neural networks have also been proposed as solutions in the 
study of 
diverse phenomena which are not as yet accessible to physical modeling~\cite{ghaboussi1991knowledge,Ghaboussi1998}. Jung and Ghaboussi~\cite{Jung2006} modeled rate-dependent materials. Stefanos and Gyan~\cite{Stefanos15} used the length of the strain trajectory traced by a material point, also called intrinsic time, as an additional input parameter in training because of its importance in
cyclic and transient loading situations. 
In the computer graphics community, deep learning technology for deformation modeling has gradually gained more attention. The DeepWarp technique~\cite{Luo2018} attempts to learn a mapping from a linear elasticity simulation to its nonlinear counterpart. Fulton et al.~\cite{Fulton2019} perform time integration of the elastodynamic system in a learned nonlinear reduced latent space, which is represented using a neural network. 
In comparison, our work focuses on the underlying constitutive model for FEM simulation and employs \rev{an} RBF \rev{representation}.

\section{Overview}\label{sec:overview}

\begin{figure*}[t!]
\centering
\includegraphics[trim=40 1.5cm 8 3.cm,clip,width=\linewidth]{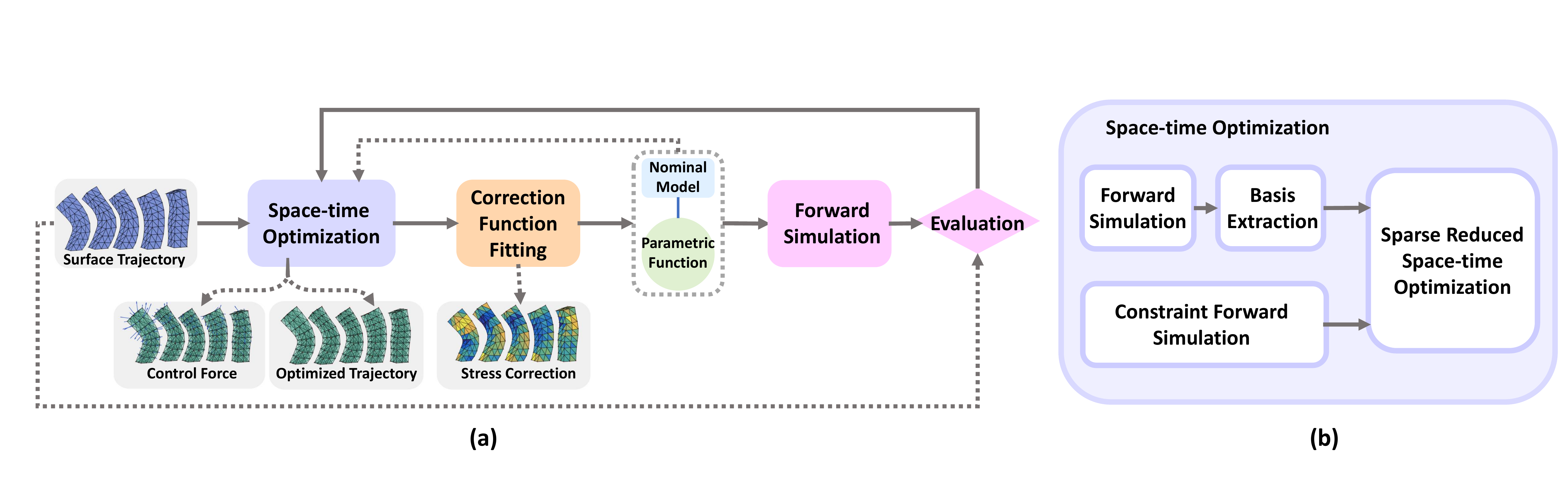}
\caption{Schematic overview of our data-driven parametric material learning framework. \urin{Our algorithm} iteratively learns a correction to a nominal material model that allows us to accurately reproduce the captured trajectory, even when the nominal model differs significantly \urin{from the corrected one}.
}
\label{fig:loop}
\vspace{-10pt}
\end{figure*}

\binTVCG{\urin{An} overview of our progressive material learning framework is \urin{given} 
in Fig.~\ref{fig:loop}(a).}
\binTVCG{Under this framework, \urin{the} material model is represented in a two component manner, \urin{as described in Section~\ref{sec:mateiral}.}} 
A baseline constitutive \urin{elastodynamic model}, called \emph{nominal model}, is assumed to be given.
The baseline model correction is encapsulated in a \emph{parametric function}.
The core of our approach is to infer \urin{this} 
parametric function
through trajectory fitting.

The input \urin{to} the system is a set of \emph{surface trajectories} of an object moving dynamically, unforced, in response to an initial perturbation. 
The main loop alternates between solving a \emph{space-time optimization problem} (Section~\ref{sec:sto}), and \emph{fitting a correction function} (Section~\ref{sec:correction}). 
\binTVCG{The space-time optimization injects a gentle \emph{control force} to keep the system trajectories close to desired input trajectories 
\urin{while still obeying} Newton's physical \pk{laws}. The key insight is that this gentle control force identifies what is currently missing due to material \urin{model} inaccuracy and should be compensated using the correction model.}
Additionally, we solve \urin{a local overdetermined algebraic} problem to identify the missing corrective stress on each tetrahedron from the vertex control forces, 
taking all the frames into consideration. Then, the best fit correction model is distilled \pk{from the strain, strain rate, and stress data}. 
Since the correction is added to both the \emph{forward simulation} and the \emph{space-time optimization}, it in turn improves how the forward simulation matches the example trajectory.
\binTVCG{ Trajectory similarity is 
monitored to determine the convergence of \rev{the} overall learning algorithm.
As the iterations progress, the correction is gradually refined to provide better accuracy. 
}


\urin{To enhance 
performance, 
we use sparse reduced space-time optimization \rev{(Section \ref{sec:spacetime})} as illustrated in Fig.~\ref{fig:loop}(b)}. 
The \emph{reduced basis is extracted} by performing singular value decomposition 
on ``snapshots'' of \urin{the} full simulation result at the beginning of each iteration using \urin{the} refined material model. Moreover, 
a \emph{constrained forward simulation} (Section~\ref{sec:warmstart}) is performed to \urin{provide} 
a good starting point for the space-time optimization \urin{procedure}, which quickly converges to a solution that identifies a plausible trajectory for unobserved nodes and the corresponding gentle control forces.

\section{Material Model}\label{sec:mateiral}

Even though there are plenty of empirical hyperelastic constitutive material models such as the nonlinear St. Venant-Kirchhoff, neo-Hookean, Ogden or Mooney-Rivlin materials, 
they do not account for all deformation phenomena that may arise. Choosing a correct model to fit measurement data is already a difficult task.  
Our data-driven approach allows us to learn a parametric
material model that can encapsulate a wide range of elastic and damping
properties in a compact and unified correction function.
%

\subsection{Nominal Material Model and Assumptions}
\label{sec:assumptions}

Our deformable models are constructed using linear shape functions.  In order to handle large deformations of soft objects, the nominal material is described in terms of the widely adopted corotated linear FEM, formulated using principal stretches~\cite{Xu2015}. \binTVCG{Corotated linear elasticity models combine the simplicity of the stress-deformation relationship in a linear material with just enough nonlinear characteristics to secure rotational invariance, and without suffering the non-physical zero stress configurations of St. Venant-Kirchhoff materials under extreme compression.} 

The deformation gradient $F$ for each tetrahedron is diagonalized by SVD, $F = U\hat{F}V^T$, and the first Piola-Kirchoff stress is computed with the principal stretches, 
$\hat{P}(\hat{F})=2\mu(\hat{F}-I)+\binPG{\zeta} tr(\hat{F}-I)I$,
where $\mu$ and \binPG{$\zeta$} are Lam\'{e} parameters.  The diagonal stress is then transformed back to the world frame, $P=U\hat{P}(\hat{F})V^T.$  
An element's contribution to its vertex forces is $P B_m$, where $B_m$ is the inverse material space shape matrix (see~\cite{Sifakis2012}).  
Summing the contributions of all elements, we can build a large sparse matrix $\mathsf{B}$, which combines the entries in $U$, $V$, and $B_m$, and can be multiplied by the block vector of all element diagonal stresses $\hat{\mathsf{p}}$ to give a block vector of \rev{all vertex forces $\mathsf{f_e}$; that is, $\mathsf{f_e}=\mathsf{B}\hat{\mathsf{p}}$.}

We include Rayleigh damping in our nominal model, with forces computed by \rev{$\mathsf{f_d}=\mathsf{D}\dot{\mathsf{x}} = (\alpha_0 \mathsf{M}+ \alpha_1 \mathsf{K})\dot{\mathsf{x}}$,
where $\mathsf{D}$ is \rev{the} 
damping matrix,} $\dot{\mathsf{x}}$ are the FEM vertex velocities, $\mathsf{M}$ is the lumped mass matrix, and $\mathsf{K}$ is the stiffness matrix assembled from per-element stiffness matrices.
The \urin{nominal model parameters are} assigned manually or computed \rev{using}
the method of Wang et al.~\cite{Wang2015}.


\subsection{Parametric Material Correction Model}
\label{sec:ParametricModel}

%

\rev{Our parametric material correction model is designed to encapsulate a correction for the elasticity and damping inaccuracies.}
Instead of explicitly adjusting 
Young's modulus and \rev{Poisson's} ratio for elasticity compensation, we directly manipulate the relationship between strain and stress. 
Similar to the work of Irving et al.~\cite{Irving2004}, the stress correction is computed in a rotated frame from diagonalized strain as \smash{$\Delta \hat{P}(\hat{F})$}, 
and then rotated back to current world frame by the same \rev{orthogonal matrices} $U$ and $V$ used to diagonalize $F$. 
The \rev{per element stress corrections}
in the world frame are
$
P_n = U \Delta \hat{P}(\hat{F})V^T.
$ 
%
%
Using principal stretches reduces the complexity of our function approximation problem, 
\rev{yet it} still permits complex stress corrections and strain dependent damping.  
Specifically, \rev{we write the strain dependent stress correction employing
an RBF with $m$ basis functions and corresponding weights $w_{k}\in\mathbb{R}^3$ as}
\rev{
\begin{equation}
\Delta\hat{P}(\hat{F}) = \sum_{k=1}^m w_{k} \phi_k( \hat{F} ).
\label{eq:rbf}
\end{equation}
}%
\rev{Therefore, all vertex elastic force corrections can be 
written as}
\rev{
\begin{equation} 
 \mathsf{\Delta }\mathsf{f_e } = \mathsf{B} \mathsf{\Delta }\hat{ \mathsf{p}},
\end{equation}
}%
\rev{in which $\mathsf{\Delta }\hat{ \mathsf{p}}$ concatenates the block vector of all element diagonal stress corrections.}

\rev{We also include an element-wise damping correction 
 through a strain dependent modification $\Delta\alpha_1(\hat{F})$ of the Rayleigh parameter $\alpha_1$. Analogous to 
Eq.~\ref{eq:rbf}, the correction
is also encapsulated in an RBF parametric representation as}
\rev{
\begin{equation}
\Delta \alpha_1(\hat{F}) = \sum_{i=k}^m w_{k}^{\alpha_1} \phi_k( \hat{F} ),
\label{eq:rbfAlpha}
\end{equation}
}%
\rev{with weights $w_{k}^{\alpha_1}\in\mathbb{R}$. Summing up the contributions from all elements, we can build a large sparse damping correction matrix $\mathsf{\Delta D}$.
The effective damping correction force is then represented as }
\rev{
\begin{equation} 
\mathsf{\Delta }\mathsf{f_d} = \mathsf{\Delta D} \,\dot{\mathsf{x}}.
\end{equation}
}%
\bin{As demonstrated in Fig.~\ref{fig:loop}(a), the weights $w$ and $w^{\alpha_1}$ are updated \pk{during each} optimization iteration.} 


\section{Force Correction Estimation} \label{sec:sto}

\begin{figure*}[t]
\centering
\includegraphics[trim=0 0.0cm 0.0cm 0.0cm,clip,width=0.98\linewidth]{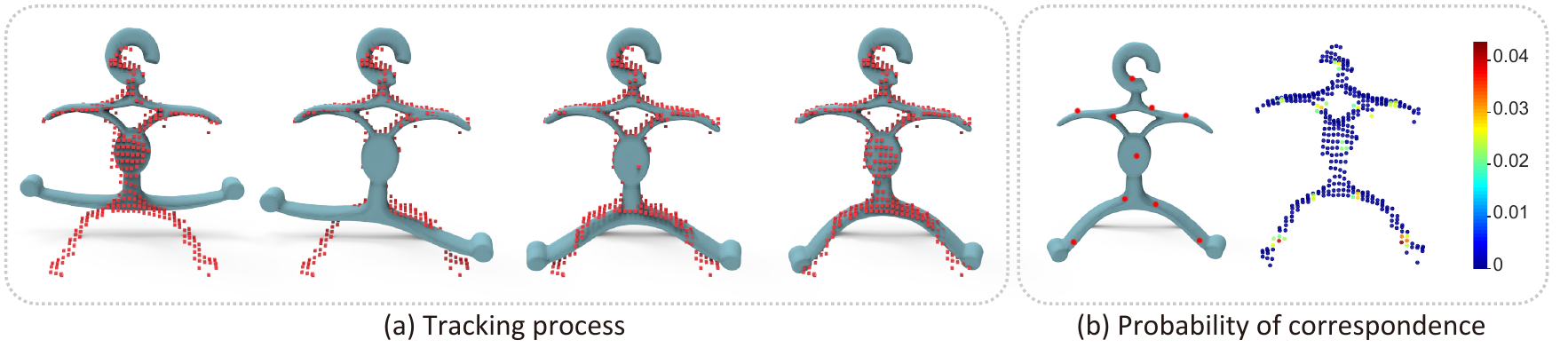}
\caption{Point cloud tracking and shape similarity matching: (a) the physics-based probabilistic tracking method gradually deforming the mesh to fit the point cloud; (b) the final maximum correspondence for each point in the point cloud to a selection of 10 surface points, which shows tight localized correspondence between the point cloud and the selected points.} 
\label{fig:HangerPointCloud}
\vspace{-10pt}
\end{figure*} 

\urin{The general philosophy of the algorithm described in this section is to gradually learn the corrections necessary to produce forward simulations that replicate the 
captured desired trajectory.}
The purpose of using space-time optimization\urin{, in turn,} is to compute a set of gentle control forces that  drive the simulation to fit 
captured data. Consequently, information to correct our currently estimated parametric material \pk{model} such that the simulation follows captured data can be distilled from \urin{the}
space-time optimization result.  

\subsection{Sparse Reduced Space-Time Optimization}
\label{sec:spacetime}

\newcommand{\norm}[1]{\left\lVert#1\right\rVert}
\newcommand{\bsfs}{\bm{\mathsf{s}}}
\newcommand{\bsfx}{\bm{\mathsf{x}}}
\newcommand{\bsfz}{\bm{\mathsf{z}}}
\newcommand{\bsfC}{\bm{\mathsf{C}}}
\newcommand{\bsfCf}{\bm{\mathsf{C_f}}}
\newcommand{\bsfCx}{\bm{\mathsf{C_x}}}
\newcommand{\bsfCz}{\bm{\mathsf{C_z}}}

Many \urin{variants} 
of the space-time constraints approach of Witkin and Kass~\cite{Witkin1988} have been proposed. To deal with the large number of degrees of freedom in our deformable models, we use reduction and sparse constraints taking inspiration from recent work~\cite{Barbic2009,Schulz2014}.
We compute a reduced basis $\Phi$ \urin{using a proper orthogonal decomposition (POD) data driven  method,} which performs principal component analysis (PCA) 
on \urin{the} forward simulation trajectory with the provided initial conditions \urin{for} our current approximation of the material model.
\footnote{An alternative would be to formulate the basis by linear combination of a much larger set of modes, and optimize the weighting matrix simultaneously~\cite{Li14}.}
We solve the space-time optimization as an error minimization problem with an
objective function that consists of two parts: \urin{physical, 
and sparse trajectory misfits}.
\urin{In practice, we optimize with reduced coordinates $\mathsf{z}_i$, where $\mathsf{x}_i = \Phi \mathsf{z}_i$. Thus,}
\uri{approximating the acceleration using a finite difference scheme}
the reduced physics errors at each time step are
\rev{
\begin{equation}
\begin{split}
\mathsf{C_f}_i \equiv \quad
& h^{-2}\Phi^T\mathsf{M}\Phi(\mathsf{z}_{i-1} - 2\mathsf{z}_i + \mathsf{z}_{i+1} ) 
- \Phi^T\mathsf{f}_{ext} \\
&-\Phi^T\mathsf{B}_{i+1} \hat{ \mathsf{p}}_{n+1} - \Phi^T\mathsf{D}_{i+1}\Phi\dot{\mathsf{z}}_{i+1} \\
& - \Phi^T\mathsf{B}_{i+1} \mathsf{\Delta} \hat{ \mathsf{p}}_{n+1}-\Phi^T\mathsf{\Delta  D}_{i+1}\Phi\dot{\mathsf{z}}_{i+1}.
\label{eq:Cfi}
\end{split}
\end{equation}
}%
This equation corresponds to our 
forward integration method because the force term 
is \urin{evaluated} at the
end of the time step.  
\urin{Next,} the desired example trajectory is sparse because it comes from an incomplete scan of the surface.
%
Letting vector $\mathsf{s}_{i}$ contain the desired point positions at time step $i$, we can write the sparse trajectory error at each time step as
\binPG{
\begin{equation}
\mathsf{C_z}_i \equiv \eta( \mathsf{S} \Phi \mathsf{z}_i - \mathsf{s}_{i} ),
\label{eq:sparsePosition}
\end{equation}}%
where the wide sparse selection matrix $\mathsf{S}$ extracts the components of the desired positions by having one non-zero entry per row.  The scalar \binPG{$\eta$} is used to specify the weight of position constraints given that the combination of physics and position constraints are solved in a soft manner.

Letting $\bsfC ( \bm{\mathsf{z}})$ concatenate all physics errors $\bsfCf$ on top of all position errors $\bsfCz$, our goal is to find a reduced trajectory $\bsfz$ that minimizes 
\urin{$\|\bm{\mathsf{C}}\|_2$}.
\rev{We solve this problem
(minimizing \scalebox{0.9}{$\frac 12 \|\bm{\mathsf{C}}\|_2^2$}) using quadratic programming, in which we need to evaluate the objective's gradient 
\scalebox{0.9}{$\left( \frac{\partial \bsfC}{\partial \bm{\mathsf{z}}} \right)^T\bm{\mathsf{C}} $}.}
%
We do not assemble the Jacobian 
matrix \scalebox{0.9}{$\frac{\partial \bsfC}{\partial \bm{\mathsf{z}}}$} directly.
Instead, we
use the chain rule and keep it in the factored form
\scalebox{0.9}{$\smash{\frac{\partial \bsfC}{\partial \bm{\mathsf{x}}} \frac{\partial \bsfx}{\partial \bsfz},}$}
where $\smash{\frac{\partial \bsfx}{\partial \bsfz}}$ simply contains copies of the basis matrix $\Phi$.
The \urin{matrix} 
$\frac{\partial \bsfCf}{\partial \bm{\mathsf{x}}}$
has a very simple part that links vertices at different time steps through the acceleration term, and a more complex part where the chain rule must be applied to compute the force gradient.  This would normally include a contribution from the parametric material correction, but generally we note better convergence when we omit it, leaving only the gradient of the nominal material. \binTVCG{The reason is \urin{that} the first order derivative of an inaccurate parametric function 
\urin{may}  introduce more \urin{noise into} 
the system. } Because we still have the parametric material correction on the right hand side, we only change the convergence and not the solution.  Thus, this second part sprinkles off diagonal terms into the matrix, linking vertices that are adjacent to a common element.  
The matrix $\frac{\partial \bsfCz}{\partial \bm{\mathsf{x}}}$
simply contains copies of the selection matrix $\mathsf{S}$.  
While the Jacobian matrix is very large, it is also very sparse.
We compute the solution using the CUSP~\cite{Cusp} library's sparse least square conjugate gradient implementation on GPU. 

We check for convergence to solution $\bm{\mathsf{z}}^*$ by monitoring our progress in reducing 
\binTVCG{errors $\bsfC ( \bm{\mathsf{z}})$.}
Once converged, \urin{the physical misfits $\bsfCf$} 
provide the necessary control forces to refine our current parametric material correction model.
That is, given an optimized reduced trajectory $\bsfz$, the gentle control force is computed as
\rev{
\begin{equation}
\begin{split}
\mathsf{f}_{i+1} =  \quad
& h^{-2}\Phi^T\mathsf{M}\Phi(\mathsf{z}_{i-1} - 2\mathsf{z}_i + \mathsf{z}_{i+1} ) 
- \Phi^T\mathsf{f}_{ext} \\
&-\Phi^T\mathsf{B}_{i+1} \hat{ \mathsf{p}}_{n+1} - \Phi^T\mathsf{D}_{i+1}\Phi\dot{\mathsf{z}}_{i+1} \\
& - \Phi^T\mathsf{B}_{i+1} \mathsf{\Delta} \hat{ \mathsf{p}}_{n+1}-\Phi^T\mathsf{\Delta  D}_{i+1}\Phi\dot{\mathsf{z}}_{i+1}.
\label{eq:physicsconstraint}
\end{split}
\end{equation}
}%
While it may be desirable to solve for control stresses at each element, as these are what is required for learning a correction, our approach permits an easier solution that directly provides a control force at each vertex.

\subsection{\binTVCG{Warm Start}}
\label{sec:warmstart}

The sparse reduced space-time optimization needs a reasonable \binTVCG{initial guess} (starting trajectory).  While the forward simulation with the current parametric material correction could serve this purpose, we find it valuable to simulate a trajectory that is also constrained to follow the desired surface motion.  
\urin{For the forward simulation, 
we solve at each step  the equation} 
\begin{equation}
\mathsf{A} \Delta \mathsf{v} = h \mathsf{f} ,
\end{equation}%
\rev{where $\mathsf{f} = \mathsf{B}\mathsf{\Delta }\hat{ \mathsf{p}} +  \mathsf{\Delta D}\dot{\mathsf{x}} +  \mathsf{f}_\text{ext}$.
Here $\mathsf{\Delta }\hat{ \mathsf{p}}$  is the block vector of stress corrections,  $\mathsf{\Delta  D}$ is the Rayleigh damping matrix correction at the current time step,
 $\mathsf{f}_\text{ext}$ is the external gravity force, and $\mathsf{A = M - hD -h^2K}$, where $\mathsf{D}$ and $\mathsf{K}$ are assembled using the
nominal model.
} Many of our models are rigidly attached to the world, and we typically remove these degrees of freedom from the system.  
We can further divide the vertices into two groups, 
\begin{equation}
\begin{pmatrix}
\mathsf{A_{uu}} & \mathsf{A_{uc}} \\ \mathsf{A_{cu}} & \mathsf{A_{cc}}
\end{pmatrix}
\begin{pmatrix}
\Delta \mathsf{v_u} \\ \Delta \mathsf{v_c}
\end{pmatrix}
=
h
\begin{pmatrix}
\mathsf{f_u} \\ \mathsf{f_c}
\end{pmatrix} ,
\end{equation}%
\rev{where we have treated the observable surface nodes as dynamic boundary constraints, and use subscripts $ \mathsf{u}$ for unconstrained and $ \mathsf{c}$ for constrained.  
Note next that $\Delta \mathsf{v_c}$ at time step $i$ can be directly computed by 
$h^{-1}(\mathsf{x}_{i-1} - 2\mathsf{x}_i + \mathsf{x}_{i+1})$, \binTVCG{in which the positions are already given as observation constraints}.}
Concequently, the second block row can be discarded\urin{, leaving} a smaller system to solve, namely,
\begin{equation}
\mathsf{A_{uu}} \Delta \mathsf{v_u} =  h\mathsf{f_u} - \mathsf{A_{uc}} \Delta \mathsf{v_c}.
\end{equation}%
\rev{Here, the effect of dynamic boundary constraints are taken into consideration on the right hand side.} 
We solve these large sparse systems using PARDISO~\cite{pardiso-6.0a,pardiso-6.0b}.


\subsection{\pk{Space-time Optimization on Noisy Real Data}}
\label{sec:realdata}
For real observations, the data can be incomplete (i.e., only partial surface scans), 
noisy, and 
in the form of unparameterized point cloud.  Thus, the exact desired point positions $s_i$ at each time step $i$ needed in Eq.~\ref{eq:sparsePosition} are no longer available. 
\binTVCG{Since now each surface tetrahedral mesh node \urin{corresponds} to a local point \urin{cloud} patch when it is close to the surface scans, 
we change the node-wise position constraints in the space time optimization to instead be patch-based position constraints.} 
We do this \urin{at} every iteration \urin{for finding $\bm{\mathsf{z}}^*$} 
in a manner 
inspired by the physics-based probabilistic tracking method proposed by Schulman et al.~\cite{Schulman13} and extended by Wang et al.~\cite{Wang2015}.

For a frame consisting of $N$ points at a given time instant (time step $i$), we denote point coordinates in the point cloud by $\bm{c}_n$  for $n = 1...N$, and node positions in the tetrahedral mesh surface by $\bm{s}_k$ for $k = 1...K$.
Let the probability of correspondence between the point cloud and the mesh nodes be $p_{kn}$.
Assuming that $\bm{c}_n$ is normally distributed around $\bm{s}_k$ as $\bm{c}_n\sim\mathcal{N}(\bm{s}_k, \Sigma_k)$ with an isotropic covariance matrix $\Sigma_k=\sigma^2\bm{I}$, then we compute the probability that nodal value $\bm{s}_k$ of the surface mesh corresponds to the observation $\bm{c}_n$ as
\begin{equation}
\mathbb{P}_{kn} = \frac{1}{\sqrt{(2\pi)^3\|\Sigma_k\|}} \exp \left(-\frac{1}{2}(\bm{c}_n-\bm{s}_k)^T\Sigma_k^{-1}(\bm{c}_n-\bm{s}_k)\right). \nonumber
\end{equation}%
\binTVCG{Note that parameter $\sigma$ is chosen to be approximately the distance between nodes on the object.}
We only assign point $\bm{c}_n$ to $\bm{s}_k$ if \urin{the probability $\mathbb{P}_{kn}$ is above a threshold}.
Fig.~\ref{fig:HangerPointCloud} shows how the mesh gradually conforms to point cloud data, along with final probabilities of correspondence between the point cloud and selected example nodes. 

Now, we can reformulate the trajectory \pk{constraints} of Eq.~\ref{eq:sparsePosition} as a weighed distance between each node and its corresponding patch of point clouds as
\begin{equation}
\mathsf{C_z}_i \equiv \binPG{\eta}  
\sum_{k,n}{\mathbb{P}_{kn}\Sigma_k^{-1}(\bm{c}_n- \Phi_k \mathsf{z}_i)} ,
\label{eq:realdata_positionconstraint}
\end{equation}%
where $\Phi_k$ gives the position of surface point $k$ from reduced coordinates $\mathsf{z}_i$, i.e., $\bm{s}_k = \Phi_k \mathsf{z}_i$.
%
The Jacobian matrix required for space-time optimization of this modified position error is straightforward to compute. 

During space-time optimization, the correspondence probability between point cloud and mesh surface is updated, which allows the mesh nodes to move freely across on the point cloud surfaces. This manner of building parameterized surface trajectories effectively uses the point cloud data as a soft constraint on the nominal model, 
and permits reasonable \uri{estimation} of surface node trajectories.

We use a very similar approach for building the initial starting point for space time optimization from real data.
The tracking procedure in Fig.~\ref{fig:HangerPointCloud} \urin{clearly demonstrates that}
the mesh can track the point cloud data correctly, even in the presence of large discrepancies.


\section{Parametric Material Correction}\label{sec:correction}

\newcommand{\Bsf}{\mathsf{B}}
\newcommand{\psfhat}{\hat{\mathsf{p}}}

\rev{This section describes how we fit the parametric material correction model defined in Section~\ref{sec:ParametricModel} using the strain and the vertex control force trajectory derived from the space-time optimization.
	
We assume to always have a variety of deformations across elements and time (i.e., the principal stretches in the data we are fitting are not all the same).  With k-means clustering, we select $m$ stretches to define 
RBFs $\phi_k(\hat{F}) = \| \hat{F} - \hat{F}_k \|$, for the cluster centers $k=1, \ldots , m$.  Then, for each time step $i$, we can assemble a tall matrix $R_i$ containing the basis functions evaluated with $\hat{F}$ for all elements, and write an equation for computing interpolated stress corrections as $\mathsf{\Delta }\hat{\mathsf{p}}_i = R_i \mathsf{w}$, 
where the unknown weights are assembled here into a block vector $\mathsf{w}$.  When damping correction is taken into consideration, the same tall matrix $R_i$ is used as well for Rayleigh correction interpolation weights $\mathsf{w}^{\alpha_1}$. 

These corrections must explain the gentle forces of Eq.~\ref{eq:physicsconstraint}, that is, 
\begin{equation}
\mathsf{f}_i  = 
\Bsf R_i \mathsf{w}
+ 
\left(\sum_j{ (R_{ij} \mathsf{w}^{\alpha_1})  K_j }\right) \dot{\mathsf{x}},
\end{equation}%
where $\mathsf{B}$ relates element stresses to nodal forces, $K_j$ is the contribution of element $j$ to the element stiffness matrix with its current strain, and $R_{ij}$ being row $j$ of matrix $R_i$. \rev{The gentle control forces may not be entirely self-consistent (i.e., an element might need different forces to correct a given state of strain at different parts of the trajectory).} Therefore, we solve for $\mathsf{w}$ and $\mathsf{w}^{\alpha_1}$ simultaneously by least squares using the data from all time steps. 
The topology independence feature of our RBF based representation  enables its
transfer to other objects.}
\section{Results and Discussion}\label{sec:results}

In 
\urin{this section,} we describe experiments that help reveal what is taking place in each step of the algorithm. 
To validate the accuracy of our algorithm, we use both \urin{real captured data and} synthetic data generated by forward simulations with known material properties.



\subsection{Space-Time Optimization}
Space-time optimization is the critical step in the entire pipeline to get training sets from pure kinematic trajectories. 
\urin{For this section}
\bin{
we designed three tests using synthetic data to illustrate \urin{how} our 
scheme can 
\urin{lead to} the convergence of the entire algorithm. To \urin{better reflect} 
real captured data in this evaluation, we use \pk{virtual scans} as input trajectories.}

\begin{figure}
\vspace{-10pt}
\includegraphics[trim=0.1cm 0.4cm 0.4cm 0.9cm,clip,width=0.9\linewidth]{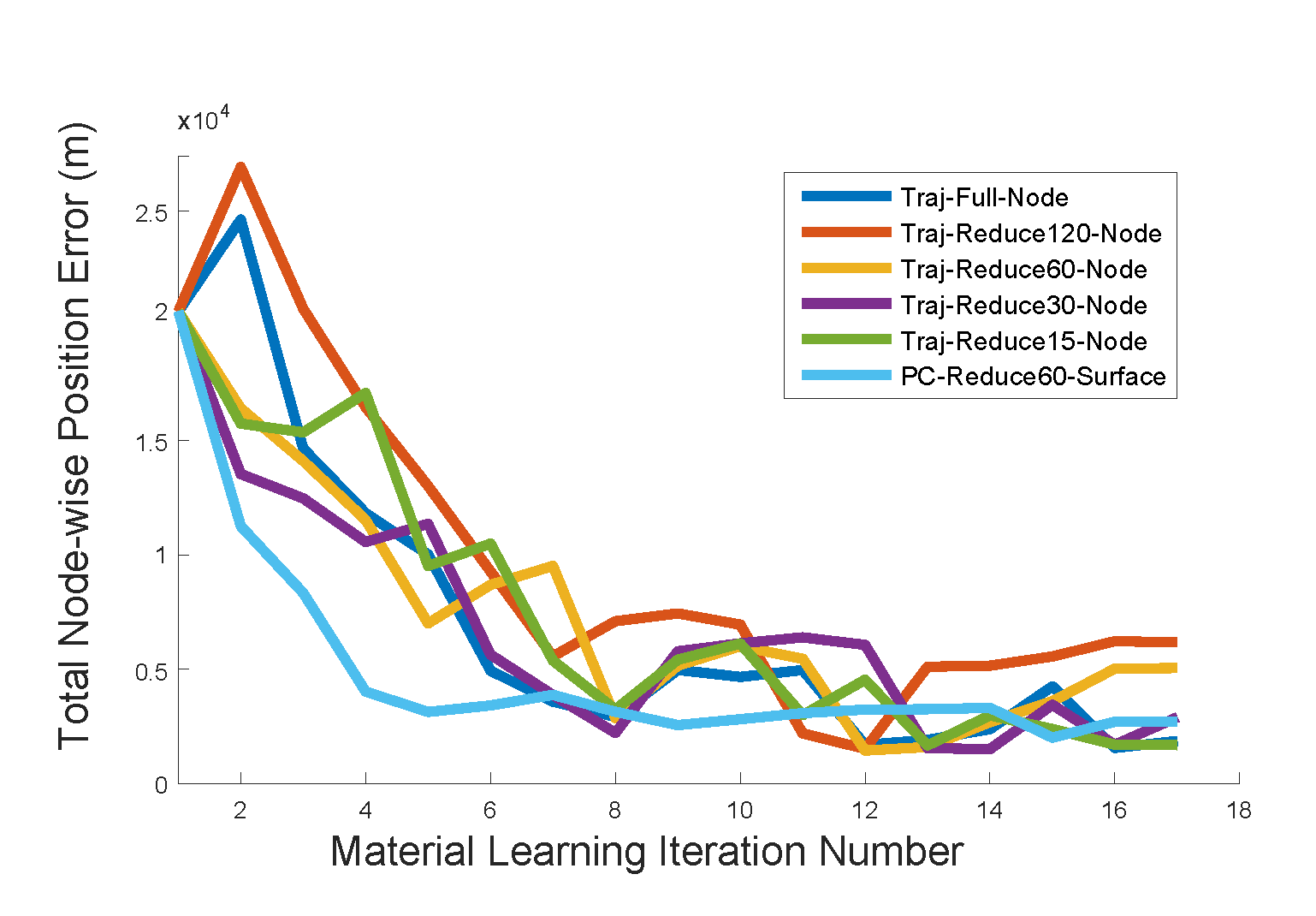}
\caption{ Convergence comparison for learning a material correction with different space-time optimization strategies. 
Input is either synthetic surface trajectories (Traj) or synthetic point cloud sequence (PC).
Full and Reduce\# distinguish between full and reduced space optimization, where \# is  the number of basis functions.}  
\label{fig:STOAccuracy}
\end{figure}

For a large scale system or a long trajectory, we must solve space-time optimization in reduced space. We test different 
strategies using the same synthetic example (turtle with neo-Hookean material) and compare how well they converge.  The results are shown in Fig.~\ref{fig:STOAccuracy}. 
%
Tests using node based position constraints or point cloud patch based position constraints are distinguished by keywords `Node' and `Surface'.  
From the comparison of the first five curves in Fig.~\ref{fig:STOAccuracy},
node-wise position constraints lead to large control forces during the first several learning iterations, when the nominal material is far from ground truth. Consequently, this introduces some overshooting. Through reduced space-time optimization, these inconveniently large force residuals are smoothly distributed throughout the entire object's domain. 
We observe good convergence as seen in the last curve in Fig.~\ref{fig:STOAccuracy} for learning from point cloud trajectories.

\begin{figure}[t]
\vspace{-5pt}
\centering
\includegraphics[trim=6cm 1cm 7cm 10cm,clip,width=.40\linewidth]{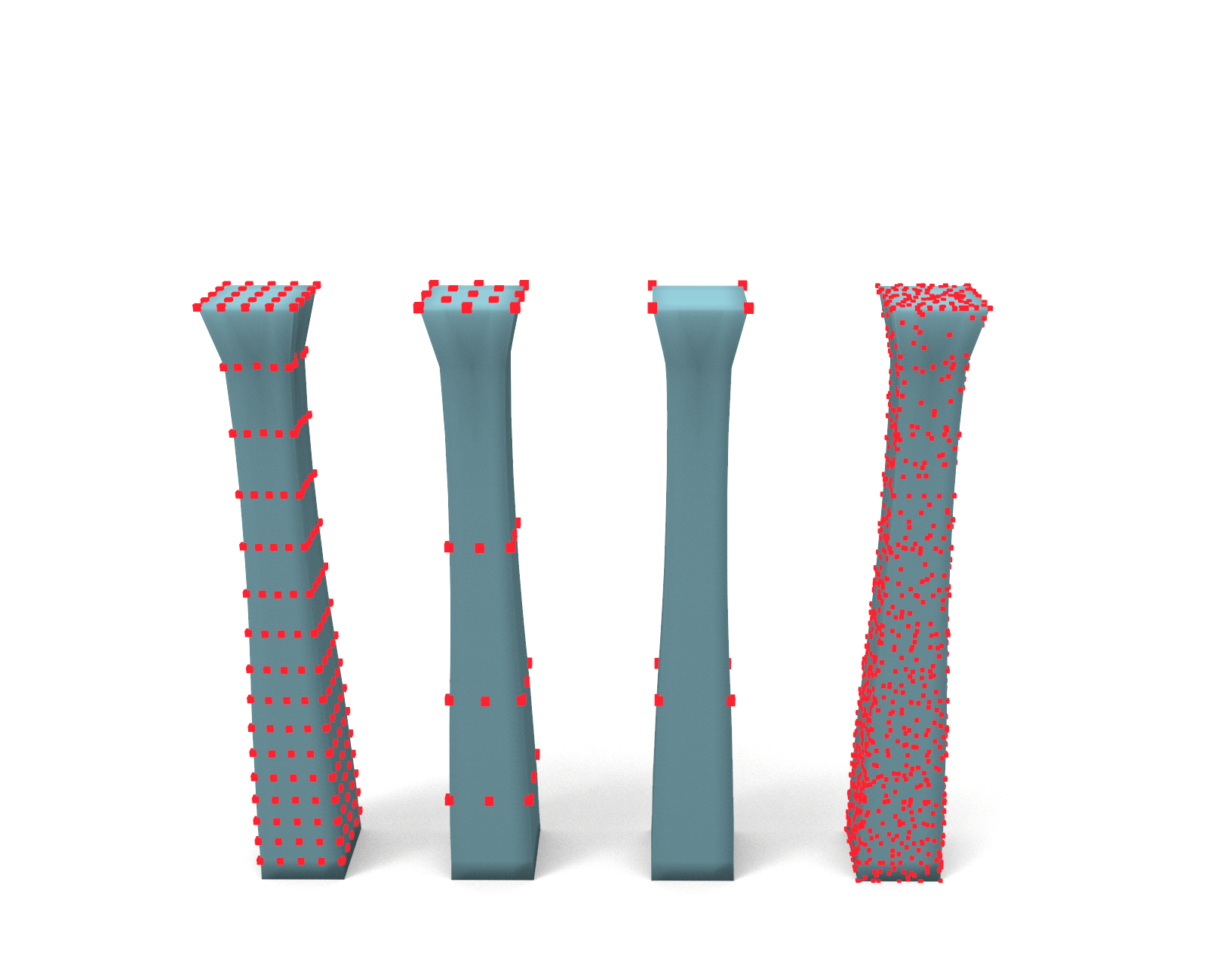}
\includegraphics[trim=0cm 0cm 0cm 1cm,clip,width=.57\linewidth]{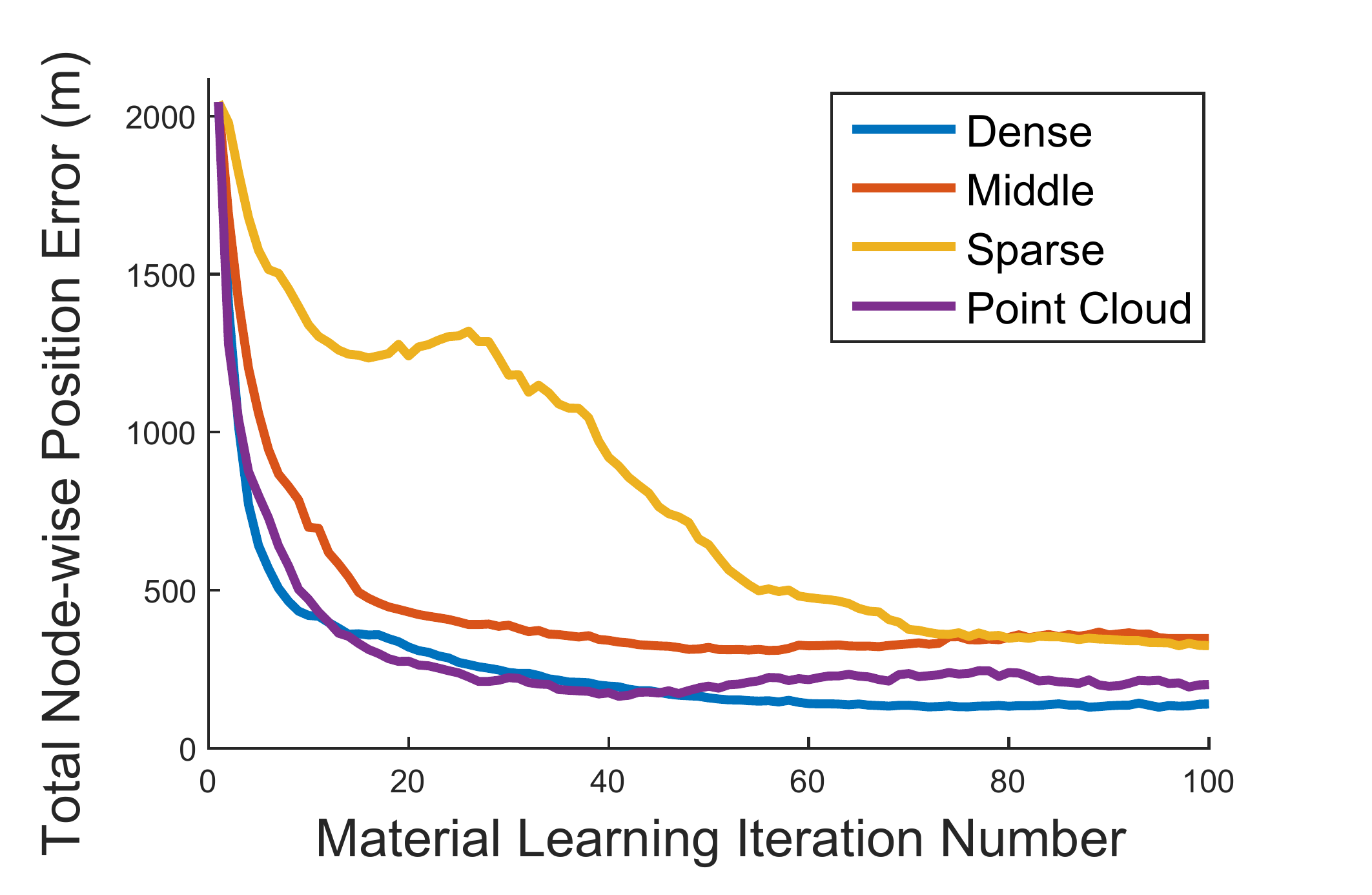}
\caption{Convergence comparison under different observation conditions. 
A neo-Hookean material with edited compression regions is the learning target. From left to right, \urin{100\%,  13\%, and 6\% of the}
surface nodes are assumed  visible. Red dots \rev{indicate} the exact location of visible nodes. 
The \urin{rightmost bar imitates a} real application where the surface information is represented as point cloud data. 
The \urin{error norms in nodal positions} 
are plotted on the right with different colors. \urin{They all converge sufficiently well.}}
\label{fig:SparseObservation}
\vspace{-5pt}
\end{figure} 

Our algorithm can handle noisy and sparse observations. We tested \urin{it on the} same synthetic examples with different observation conditions. 
{The learning target is a} neo-Hookean material with edited compression regions.
\urin{Notice that in} Fig.~\ref{fig:SparseObservation}, the \pk{number of observation} points in the first three cases drops sharply.
For all these experiments, our algorithm can converge to a \urin{sufficiently accurate} solution. 
We also used virtual scan\pk{s} to imitate a real capture situation, where point cloud data cover the whole surface. Our patch based position constraint adopted in the last case (Section~\ref{sec:realdata}) performs better than the sparse observation case and only mildly worse than the accurate full surface observation case. 

\newcommand{\tabincell}[2]{\begin{tabular}{@{}#1@{}}#2\end{tabular}}
\renewcommand{\arraystretch}{1.0} 
\setlength{\tabcolsep}{5.5pt}
\begin{table*}[t]
\begin{center}
\caption{
Parameters and errors for different test cases.  For each test object we list ground truth (G) and nominal (N) parameters
for Young's modulus $E$ in MPa, Poisson ratio $\nu$, Rayleigh damping $\alpha_{0}$ and $\alpha_{1}$, and use U and R to denote user designed and real materials, respectively.  The maximum position errors are measured using percentage of object size in simulations using a time step of 0.001 seconds.}
%
%
\label{tab:test_case_statistic}
\small
\begin{tabular}{|l|l|c|c|c|c|c|c|c|c|c|c|c|}
\hline
Case	 &Material, Ground Truth (G) & $E_G$ & $\nu_G$ & $\alpha_{0G}$ & $\alpha_{1G}$ &Material (N) & $E_N$ & $\nu_N$ & $\alpha_{0N}$ & $\alpha_{1N}$ 	&${err}_L$ \\
\hline
Turtle		&neo-Hookean   						& 2e4  		&0.43  	&0.0   	&0.0   &Corotation & 3.5e4  & 0.43  &0.0   &0.0				& 2.8 \\
Dragon 			&StVK		  						& 1e5 		&0.40  	&0.0   	&0.0   &Corotation & 1.2e5  & 0.40	&0.0   &0.0			& 0.4 \\
Sphere1 			&
neo-Hookean(tension)
& U  		&U  	&0 		&0.2   &Corotation & 1.2e5  & 0.43  &0.0   &0.2		& 4.1\\
Sphere2 			&
neo-Hookean (compression)
& U  		&U  	&0  	&0.2   &Corotation & 1.2e5  & 0.43  &0.0   &0.2		& 6.0\\
Bar (Damping1)
& 
Corotation + strain-dep damping
& 4e4  		&0.43   &U   	&0.2   &Corotation & 4e4 	& 0.43  &0.0   &0.2		& 0.1\\
Bar (Damping2)	
&
Corotation + strain-dep damping
& 4e4  		&0.43   &U   	&0.2   &Corotation & 4e4 	& 0.43  &0.0   &0.2			& 0.8\\
Pot Holder	&Real Material	  					& R  		&R   	&R   	&R     &Corotation & 2.4e6  & 0.43  &0.0   &0.0			& 3.0/7.0\\
\binnew{Hanger}			&Real Material	  					& R  		&R   	&R   	&R     &Corotation & 4.5e5	& 0.43  &0.001   &1.0			& 4.0\\
\binnew{Silicon Bar}			&Real Material	  					& R  		&R   	&R   	&R     &Corotation & 3.0e5	& 0.45  &0   &0.0			& 6.0\\
Bar (Heterogeneous)
&Corotation	  						& 1e5/1e7 	&0.40  	&0.0   	&0.0   &Corotation & 3e6    & 0.40  &0.0   &0.0	 	& 0.2/0.5 \\
\hline
\end{tabular}
\end{center}
\vspace{-10pt}
\end{table*}

\begin{figure}[t]
\vspace{-10pt}
\includegraphics[trim=3cm 2.5cm 1cm 1.5cm,clip,width=.99\linewidth]{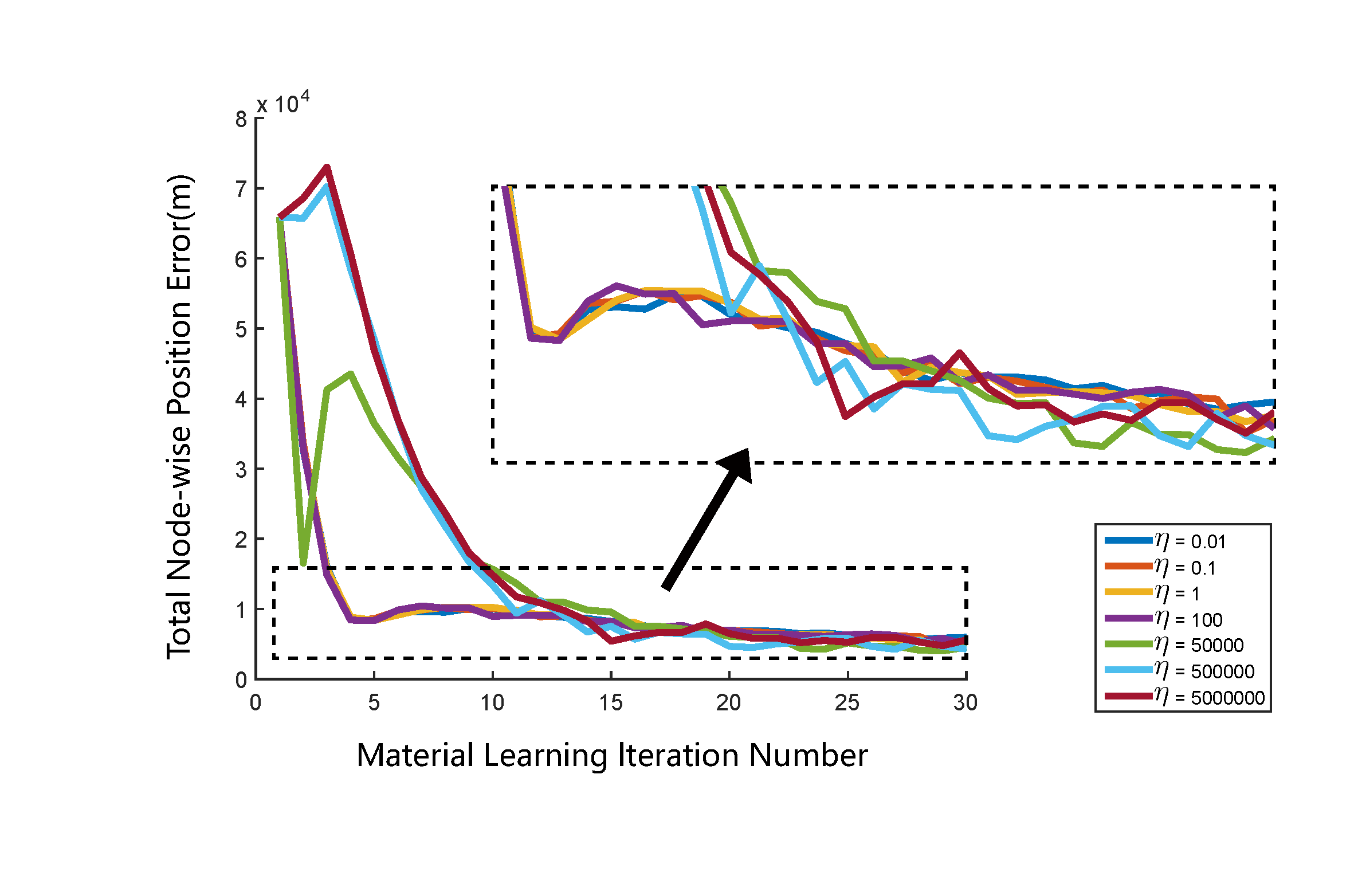}
\caption{Convergence comparison with different position constraint enforcement weights. 	
The learning target is a neo-Hookean material with edited compression regions.  
Position error norms at each iteration are plotted for seven different $\lambda$ \urin{values} in Eq.~\ref{eq:sparsePosition}. Strong position constraints introduce vibration at early iterations,
while soft ones mildly sacrifice accuracy.  }
\label{fig:LambdaConvergy}
\vspace{-15pt}
\end{figure} 
\bin{
As \urin{described in \pk{Eq.~\ref{eq:sparsePosition}}}, $\lambda$ controls the enforcement \urin{weight} 
of position constraints, which influences the control force. 
We chose different $\lambda$ \urin{values in the wide range} [0.01, 500000], and \urin{simulated the} 
same example (turtle with neo-Hookean material as learning target, \pk{all} surface nodes assumed visible). 
From Fig.~\ref{fig:LambdaConvergy}, we observe that larger $\lambda$ \urin{values} introduce some vibration or even overshooting at 
the \urin{early optimization stages}.
The unbalanced ratio between force residual constraint and position constraint \urin{causes a large force deviation in} the internal points, 
which consequently decreases the quality of generated training data.
\urin{In contrast, the curves corresponding to} smaller $\lambda$ are much smoother along \urin{the optimization iteration axis}. 
From a global perspective, different~$\lambda$ \urin{will not necessarily}  produce significantly different final results, even though there are still subtle accuracy 
\urin{drops} for smaller $\lambda$ cases as illustrated in \urin{the} zoom-in of Fig.~\ref{fig:LambdaConvergy}. 
}
\bincheck{\pk{We expect it should be possible to} adaptively tune $\lambda$ \pk{between} iterations to speed up convergence. 
For real captured data, $\lambda$ is chosen to be small, \urin{depending on the confidence in the observations}. }

\begin{figure}[t]
\includegraphics[trim=2cm 1cm 2cm 1.5cm,width=.96\linewidth]{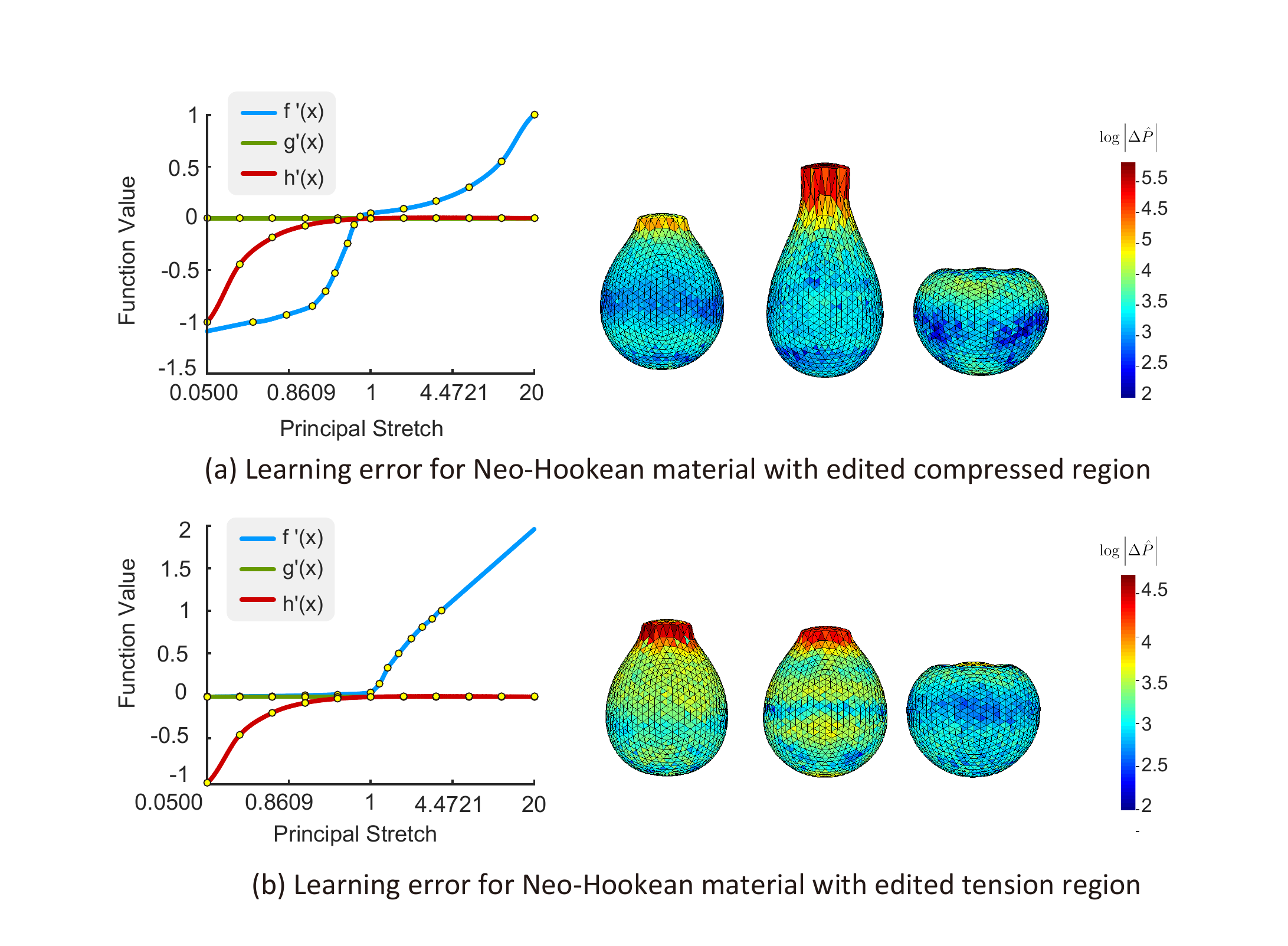}
\caption{ Visualization of the learning result accuracy for user designed elastic material examples~\cite{Xu2015}. 
Plots (left) show edited compression and tension regions of a Neo-Hookean material.  Images (right) use color to show principal stress error distribution of our learned material correction to a corotational model.}
\label{fig:userdesigenelasticity}
\end{figure}

\subsection{Nonlinear Constitutive Material Modeling}
To validate the generality of our parametric material model estimation algorithm, we test its ability to learn a variety of different materials using a corotational model for the nominal material.  Ground truth trajectories are either generated in  the VEGAFem~\cite{Vega} library using classical hyperelastic material model, 
or \urin{they are} user defined elasticity and damping models, or captured by Kinect sensors.  Table~\ref{tab:test_case_statistic} shows the statistics of all our testing cases.  Qualitative results for these cases can be seen in the supplementary video.  Each case is discussed below, while reconstruction errors are listed in the last column of  Table~\ref{tab:test_case_statistic}.

\subsubsection{Classical Hyperelastic Material}
\urin{In Fig.~\ref{fig:tease}, the turtle 
is made of neo-Hookean material, and} the dragon is made of StVK material. 
We use two deliberately designed test trajectories to validate our learning result. The first test has a similar deformation scale as the training trajectory, 
while the second test has a much larger range of deformation. Table~\ref{tab:test_case_statistic} and the supplementary video show that the learning result can reproduce similar deformation with high accuracy; the results for different deformations also demonstrate low error. Vibration differences can only be observed towards the end of a sequence, and as such, can largely be attributed to error accumulation. 

\subsubsection{User Designed Elasticity and Damping Model}

\begin{figure}[t]
\vspace{0pt}
\includegraphics[trim=3cm 7.5cm 1cm 1.5cm,clip,width=.89\linewidth]{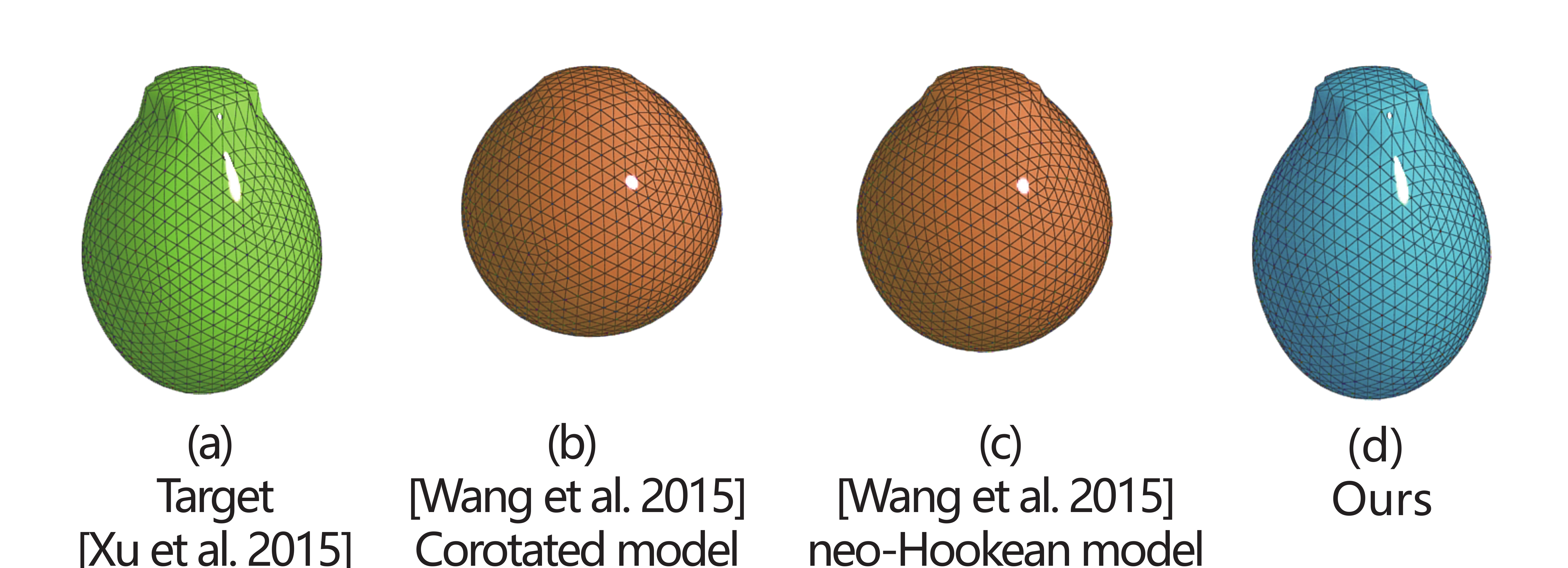}
\caption{Comparing material learning techniques where (a) the target trajectory is generated using neo-Hookean material with modified compression region~\cite{Xu2015}; 
(b)~\cite{Wang2015}; (c) \cite{Wang2015}, replacing 
the corotated model with the neo-Hookean model; and (d) our algorithm. }
\label{fig:ComparewithWang}
\end{figure}

Our algorithm can \urin{also be} extended to model user designed nonlinear elasticity and damping material models. 
Our third example is a soft solid sphere whose top part is keyframed in an up-down motion (Fig.~\ref{fig:userdesigenelasticity}). Following the method of Xu et al.~\cite{Xu2015}, the internal elastic forces and tangent stiffness matrices are formulated in a polynomial space of principal stretch.  Customized materials are designed by editing a single stress-strain curve using a spline interface. We tested two different designed nonlinear materials which edit tension and compression starting from the a neo-Hookean material as shown in Fig.~\ref{fig:userdesigenelasticity}. The supplementary video shows indistinguishable simulation trajectories, while Fig.~\ref{fig:userdesigenelasticity} shows small stress reconstruction errors produced by our parametric material correction of a nominal corotational model.
\bin{We also compared with \urin{the} parameter fitting based method of \pk{Wang et al.~\cite{Wang2015}} and 
\pk{a modification of the method}
which \urin{replaces} the simple corotated elasticity model with a neo-Hookean model. The \urin{reported} side by side comparison in Fig.~\ref{fig:ComparewithWang}
and the supplementary video both \uri{clearly} demonstrate \urin{that} our method is superior to \urin{this}
parameter fitting algorithm, especially when the default model is simple.
}

\begin{figure}[t]
\includegraphics[trim=0cm 0cm 0cm 0.8cm,clip,width=\linewidth]{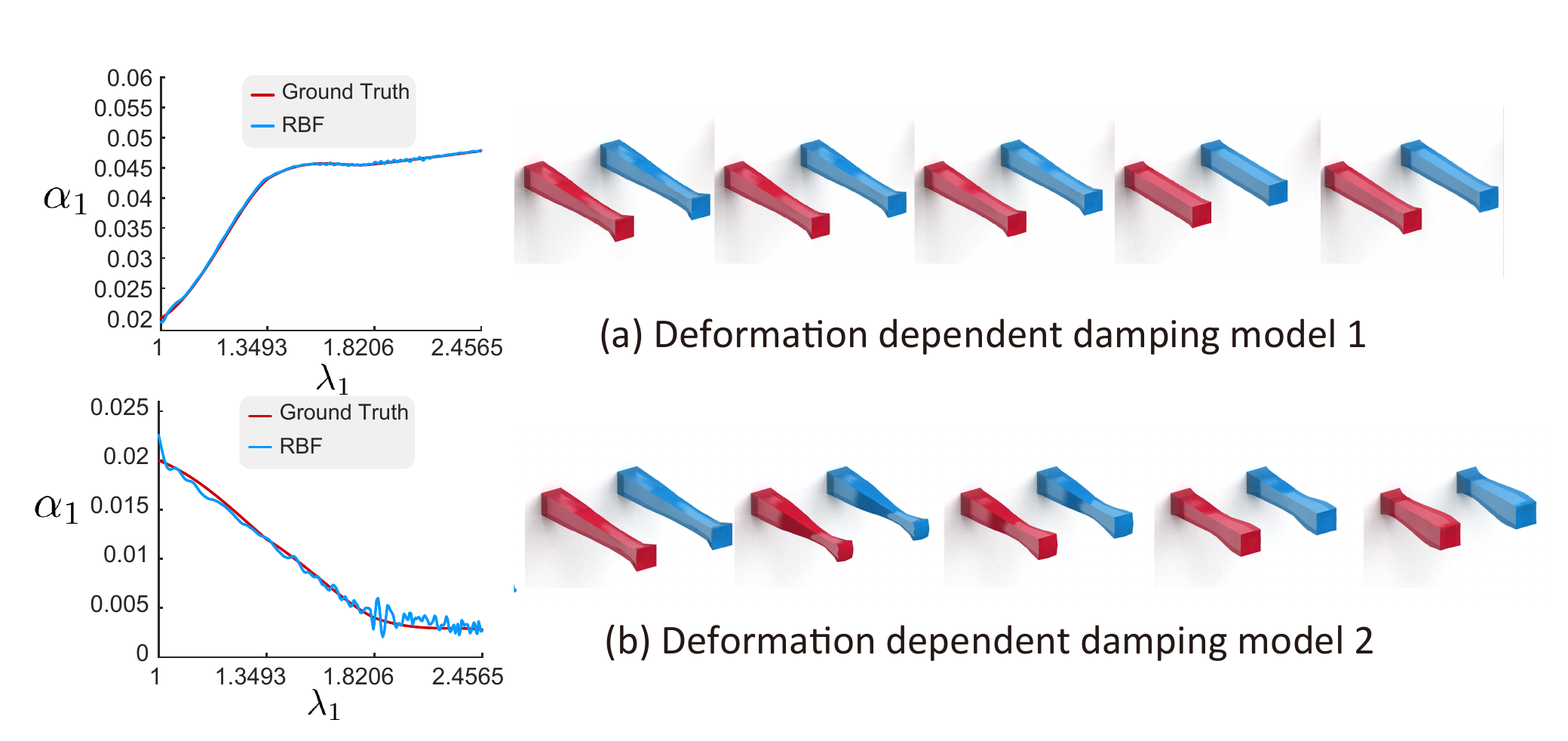}
\caption{ Visualization of the learning result accuracy for user designed deformation dependent damping models. Red is used to show ground truth, while blue shows simulated poses with a learned parametric material correction. }
\label{fig:userdesigndamping}
\vspace{-5pt}
\end{figure} 

\begin{figure*}[tbh]
\centering
\includegraphics[trim=14cm 2cm 17cm 2cm,clip, width=0.12\textwidth]{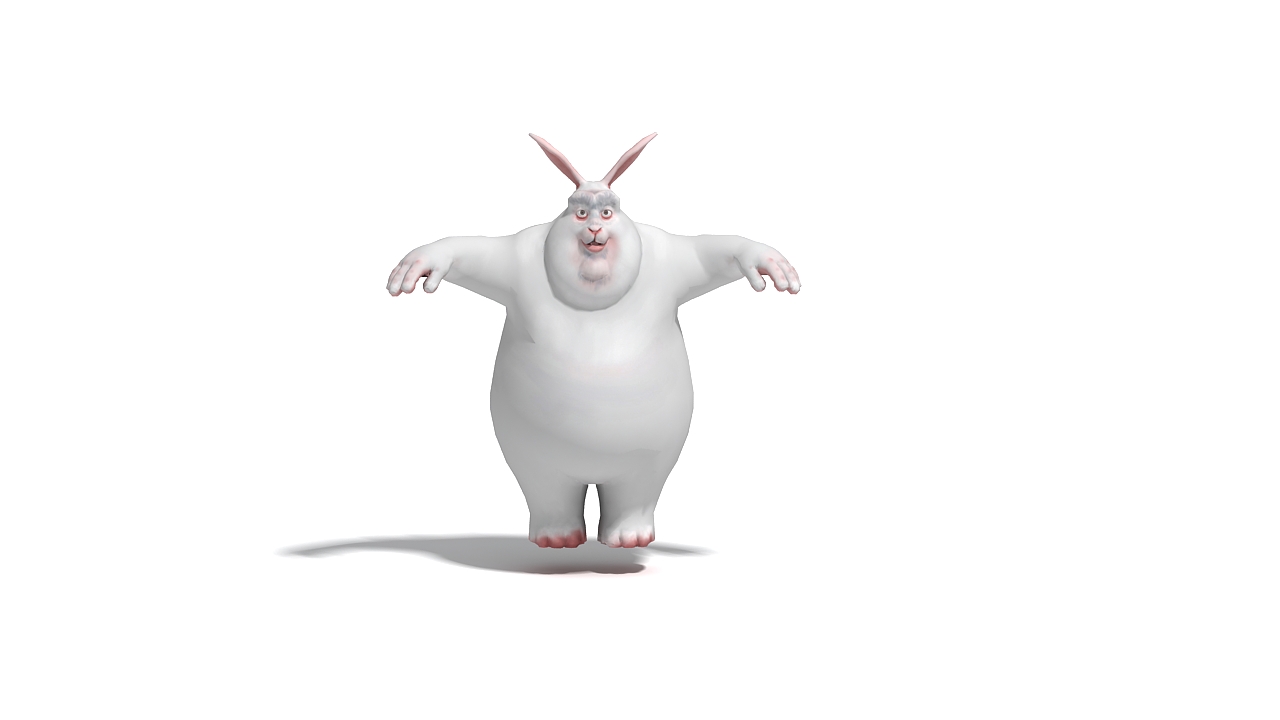}
\includegraphics[trim=14cm 2cm 17cm 2cm,clip,width=0.12\textwidth]{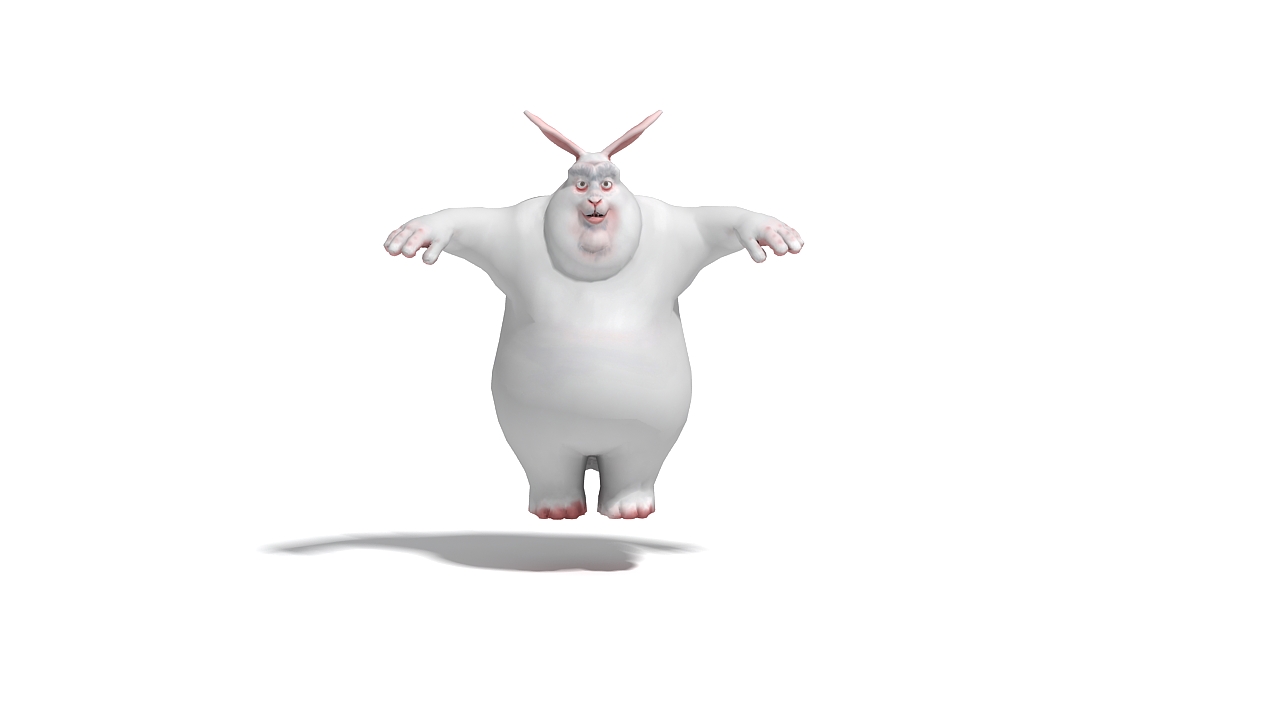}
\includegraphics[trim=14cm 2cm 17cm 2cm,clip,width=0.12\textwidth]{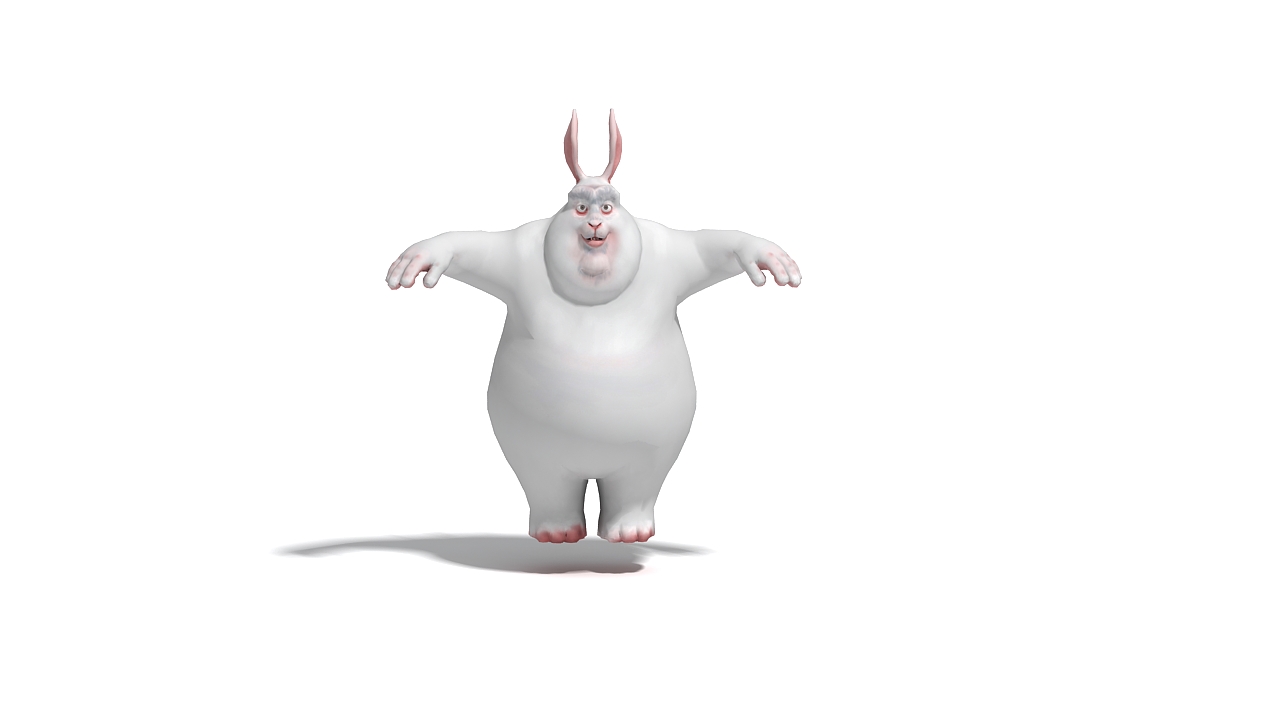}
\includegraphics[trim=14cm 2cm 17cm 2cm,clip,width=0.12\textwidth]{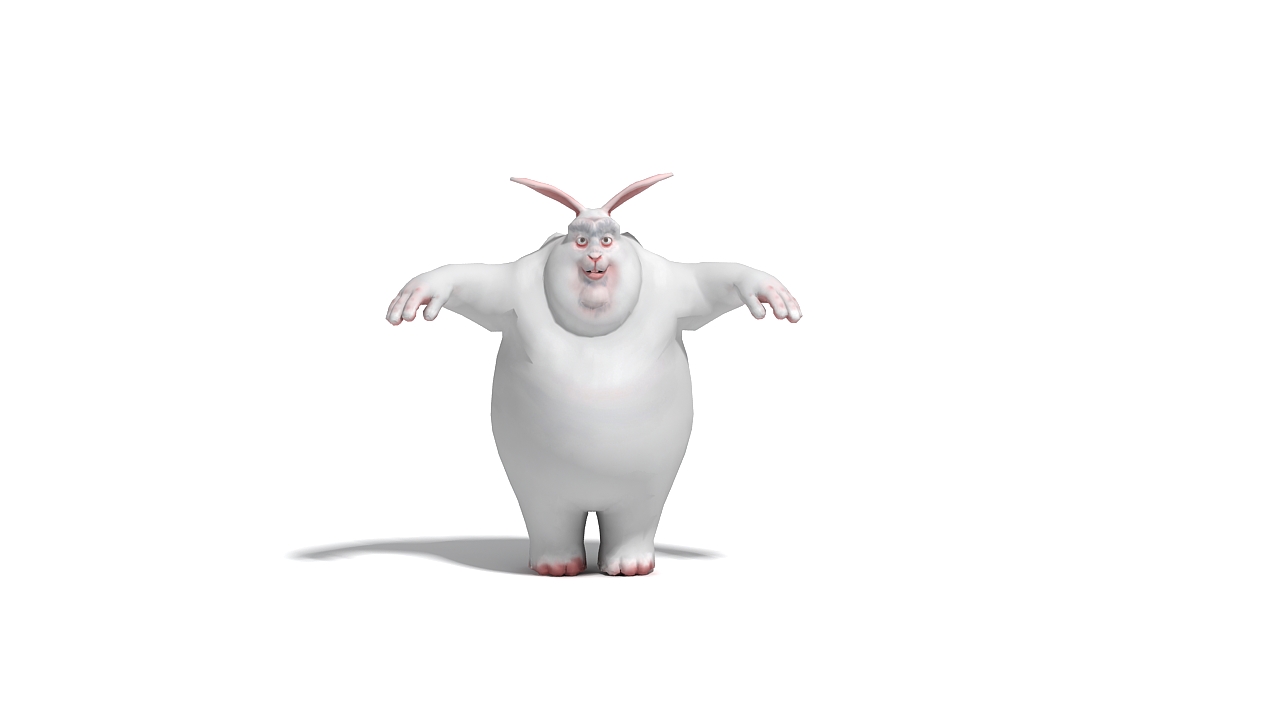}
\includegraphics[trim=14cm 2cm 17cm 2cm,clip,width=0.12\textwidth]{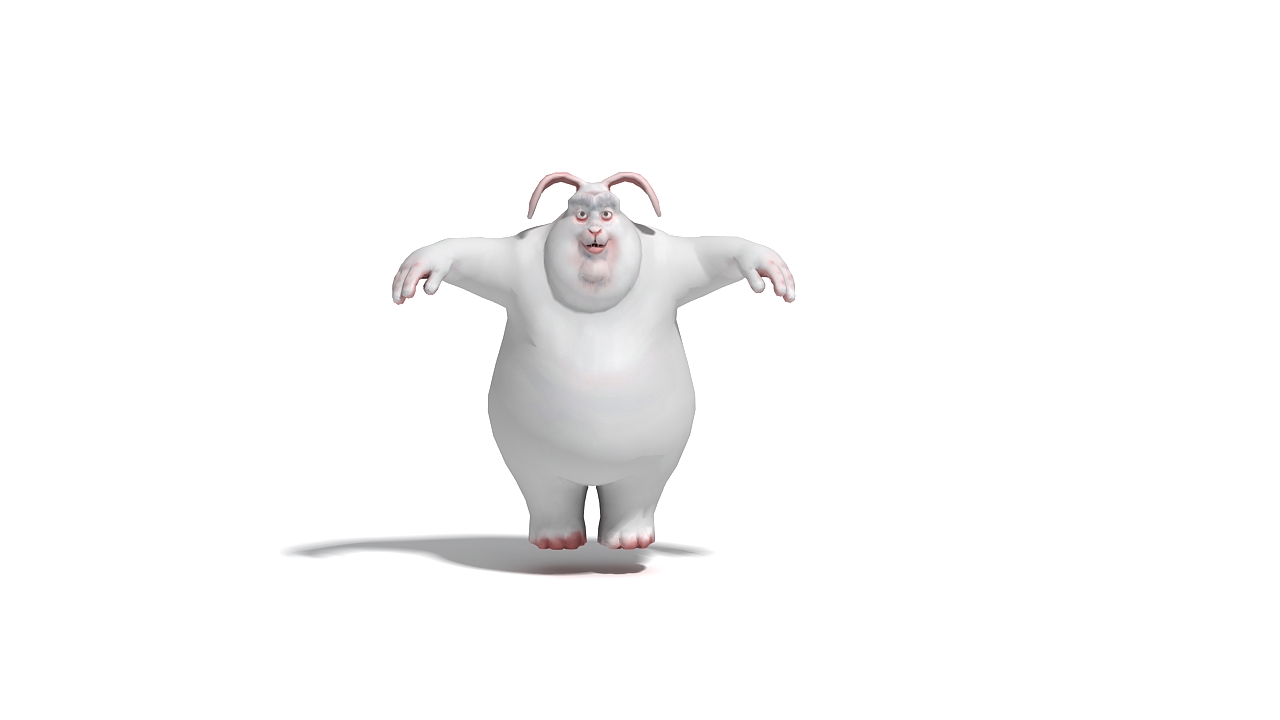}
\includegraphics[trim=14cm 2cm 17cm 2cm,clip,width=0.12\textwidth]{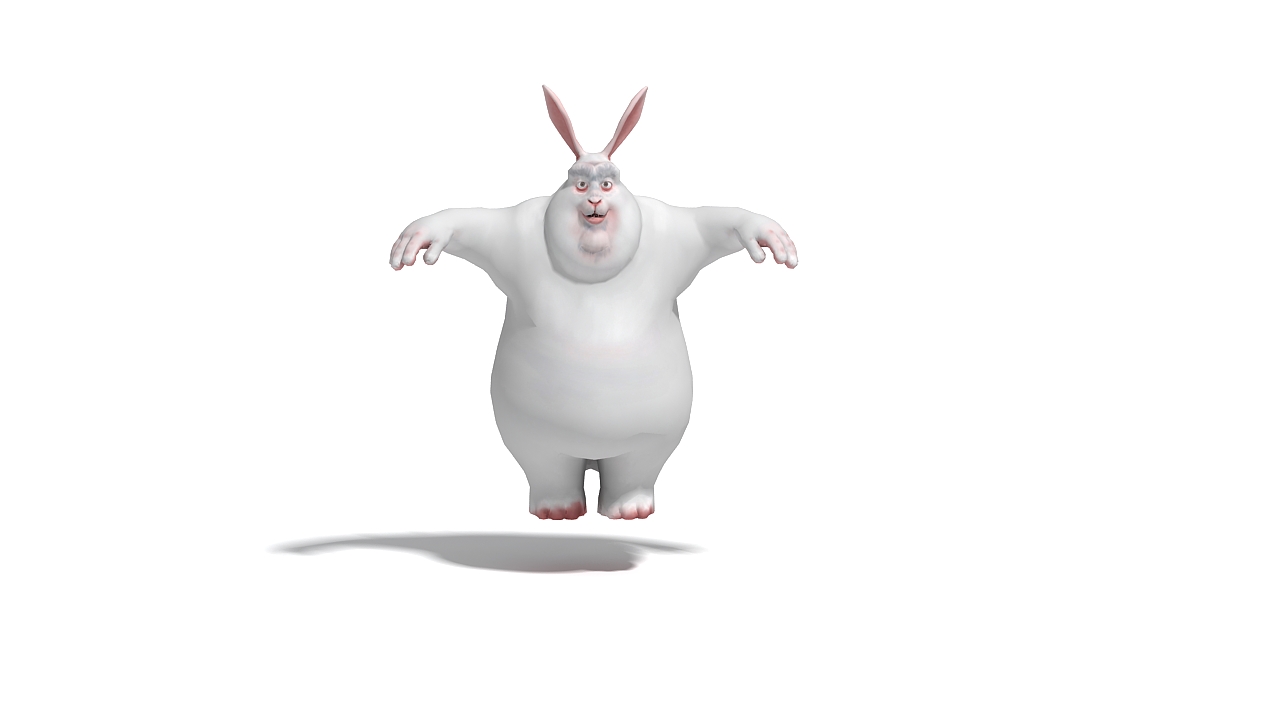}
\includegraphics[trim=14cm 2cm 17cm 2cm,clip,width=0.12\textwidth]{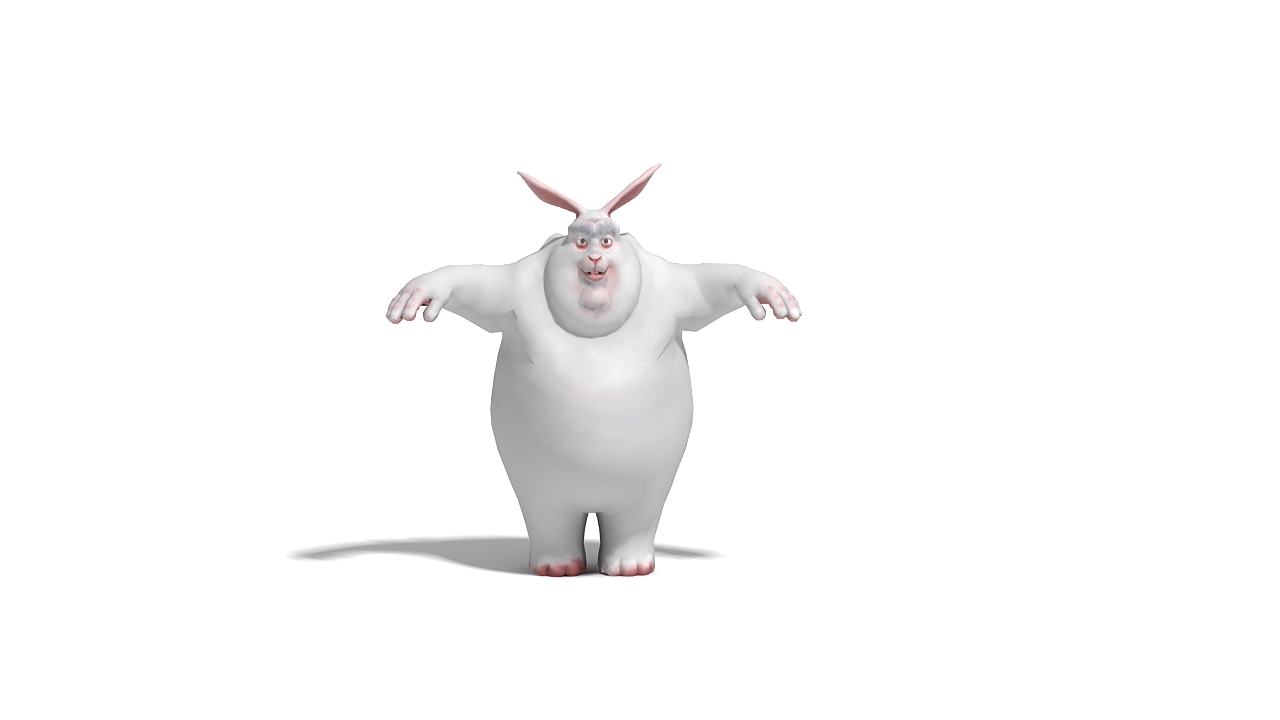}
\includegraphics[trim=14cm 2cm 17cm 2cm,clip,width=0.12\textwidth]{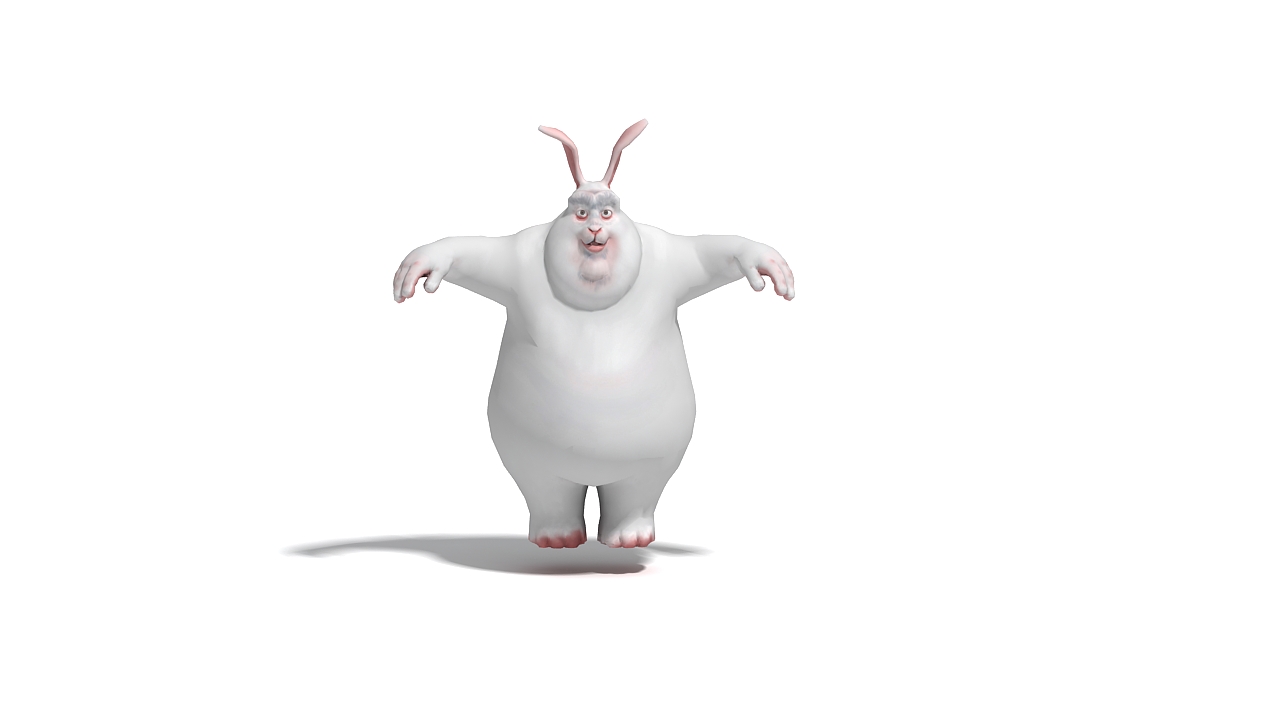}
\\
\includegraphics[trim=12cm 2cm 12cm 2cm,clip,width=0.12\textwidth]{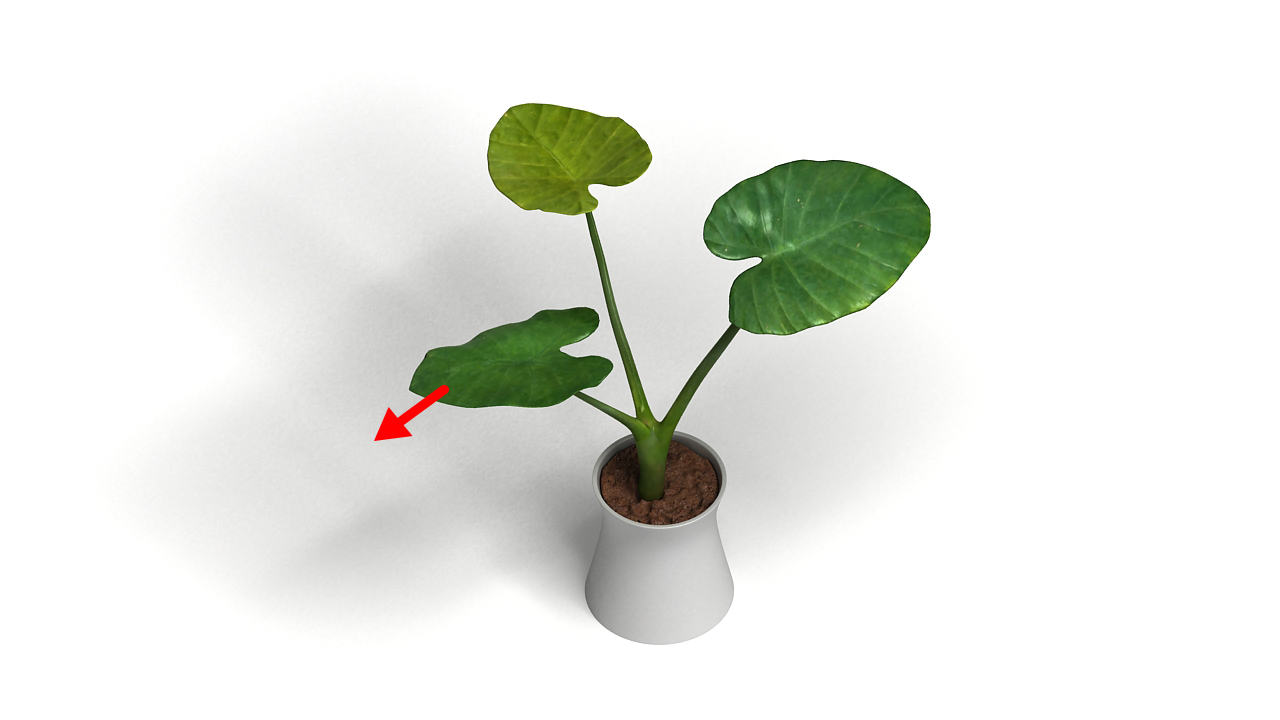}
\includegraphics[trim=12cm 2cm 12cm 2cm,clip,width=0.12\textwidth]{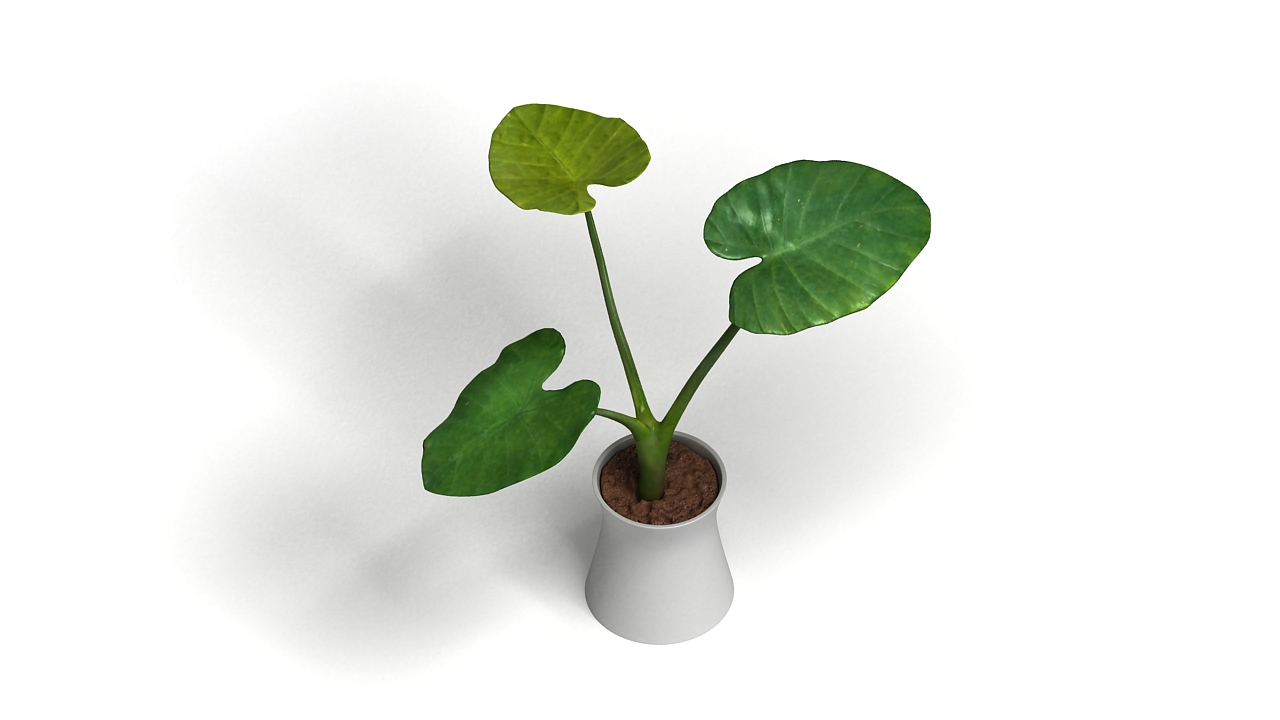}
\includegraphics[trim=12cm 2cm 12cm 2cm,clip,width=0.12\textwidth]{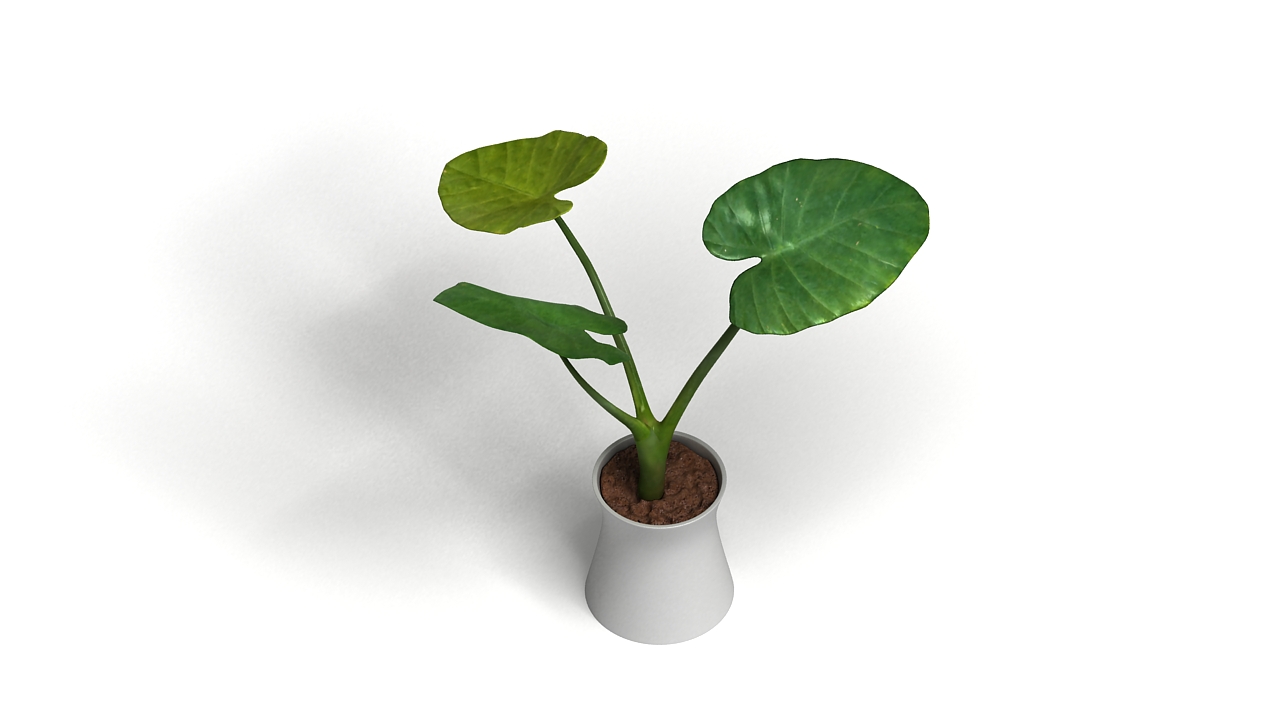}
\includegraphics[trim=12cm 2cm 12cm 2cm,clip,width=0.12\textwidth]{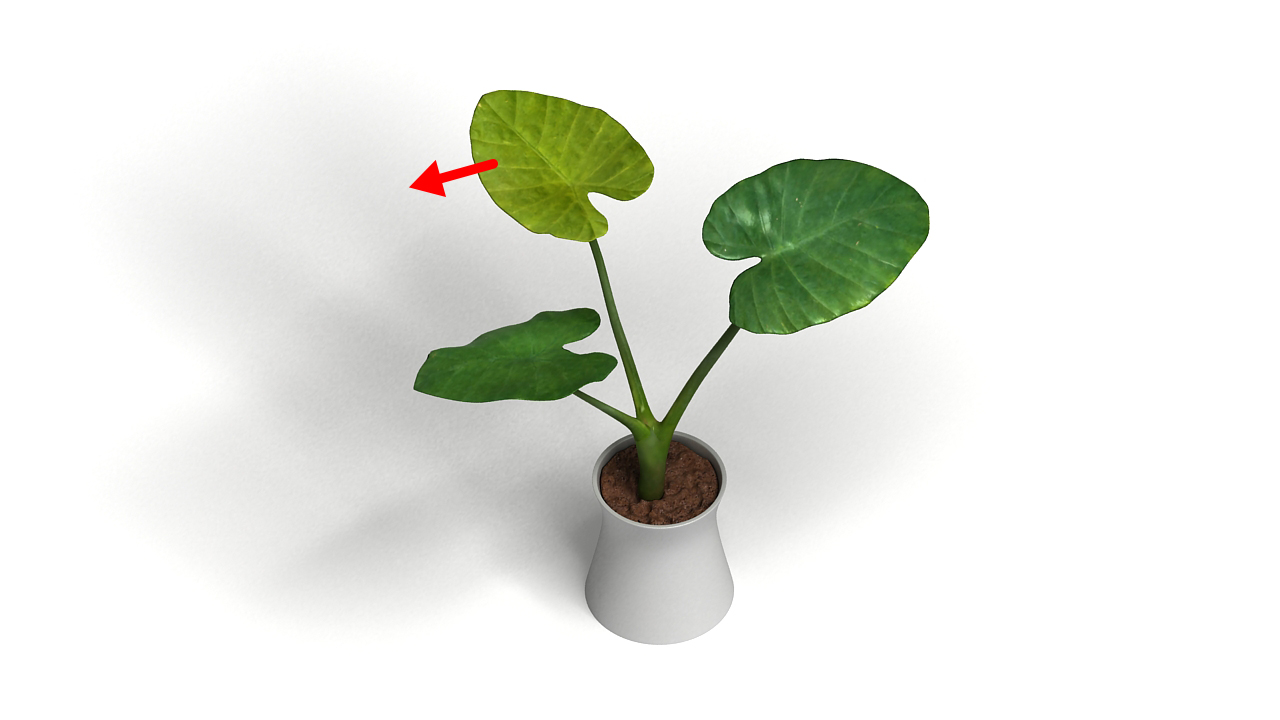}
\includegraphics[trim=12cm 2cm 12cm 2cm,clip,width=0.12\textwidth]{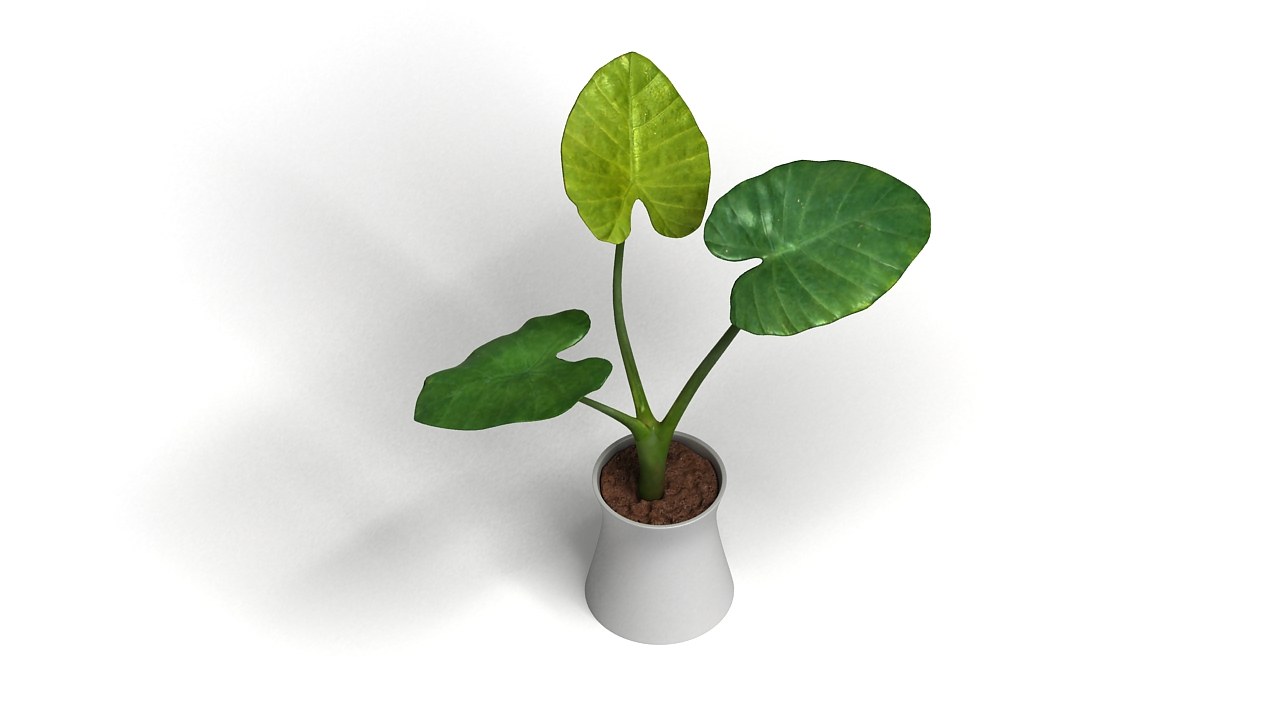}
\includegraphics[trim=12cm 2cm 12cm 2cm,clip,width=0.12\textwidth]{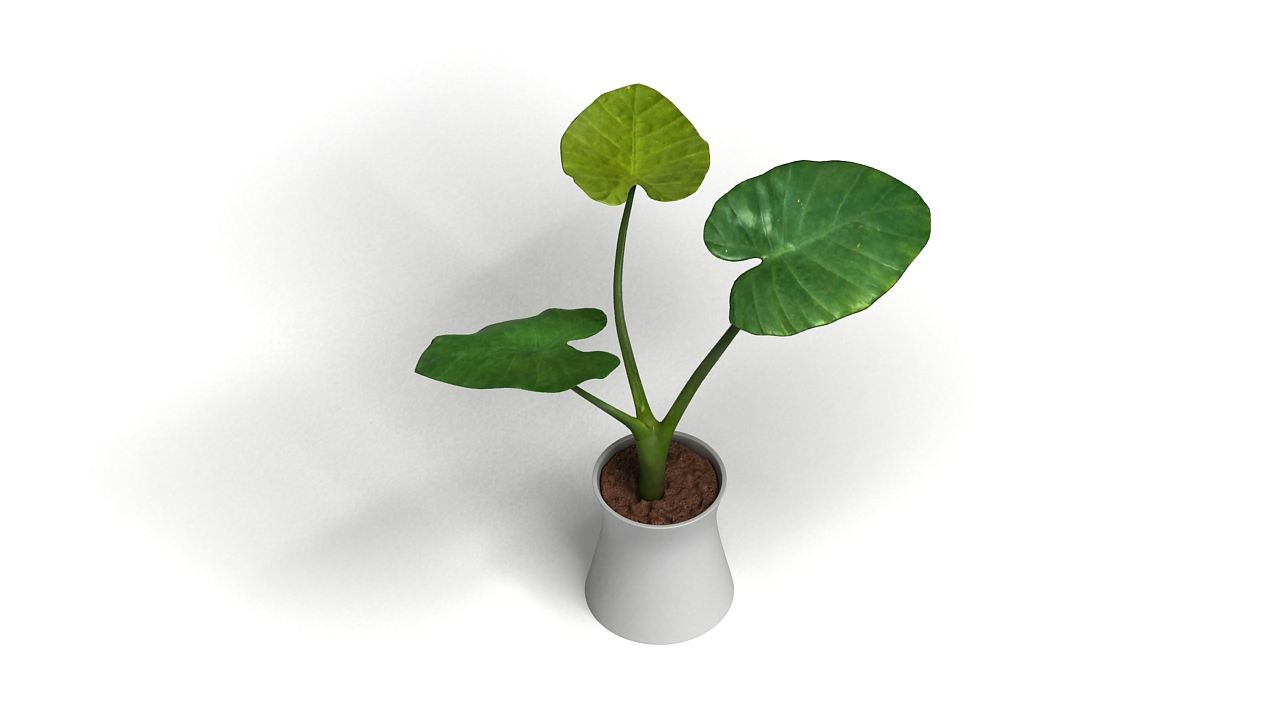}
\includegraphics[trim=12cm 2cm 12cm 2cm,clip,width=0.12\textwidth]{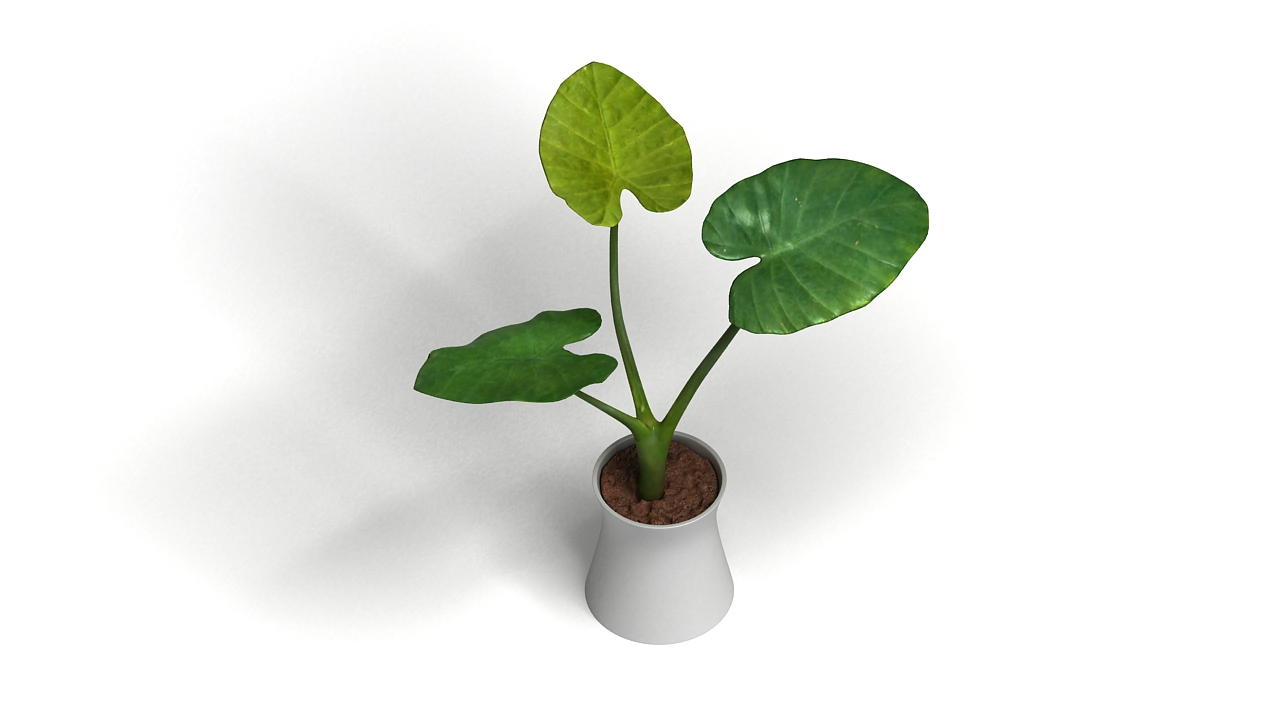}
\includegraphics[trim=12cm 2cm 12cm 2cm,clip,width=0.12\textwidth]{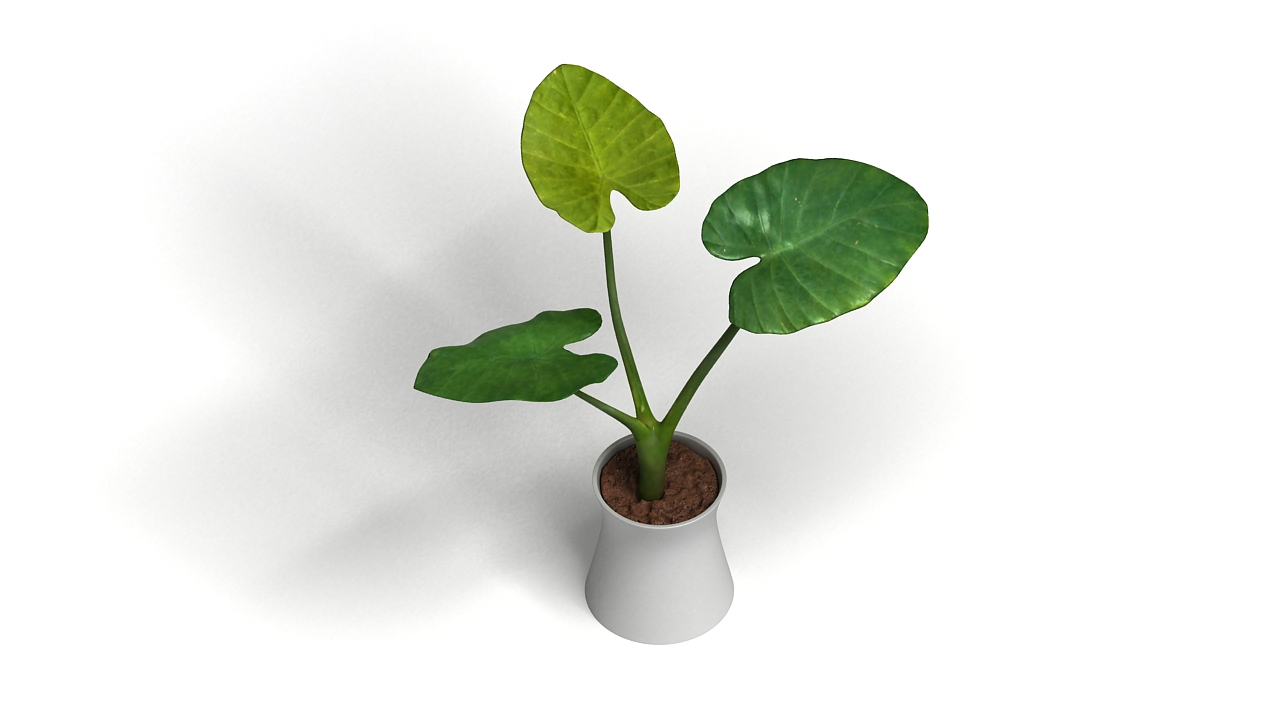}
\caption{Material transfer. Top: A chubby jumping bunny uses the learned material of the ball in Fig~\ref{fig:userdesigenelasticity}(b).
Bottom: A plant responds to user interactions, where the left and middle leaves use learned damping properties of the bars in Fig.~\ref{fig:userdesigndamping}.}
\label{fig:synthesis}
\vspace{-10pt}
\end{figure*} 

Although linear viscous damping \pk{is} widely used in the computer graphics community, \pk{this} only constitutes a small subset of all viscous damping models. Under many circumstances, the damping matrix $C$ can depend nonlinearly on both deformation and velocity. To validate the accommodation of our material model estimation algorithm for damping compensation, Fig.~\ref{fig:userdesigndamping}
shows our tests on a strain-dependent damping model. We start from a Rayleigh damping model, and substitute the original constant stiffness damping coefficient $\alpha_1$ with a polynomial function of the first principal stretch $\lambda_1$. The function of deformation dependent $\alpha_1$ is controlled using a spline tool.
%
%

Since our parametric material correction model is independent of topology, it can be easily transferred to other simulation scenarios. The neo-Hookean material with modified tension region model which we learned from a ball example
can be transferred, for \urin{instance}, 
to a chubby bunny. As \urin{can be gleaned from the}
video and \rev{Fig.~\ref{fig:synthesis} (top)}, the bunny belly vibrates \urin{in a lively fashion} during jumps. 
\rev{From a side by side comparison with the ground truth result in the video, the simulated trajectories match well initially but then drift apart due to error accumulation. This inaccuracy could be due to the learned material itself and/or the mismatch in the geometric model.} We also transfer the two deformation dependent damping models of Fig.~\ref{fig:userdesigndamping} to different leaves of a taro plant to assign separate properties for young and old leaves as shown in the video and Fig.~\ref{fig:synthesis}
\rev{(bottom)}.

\subsubsection{Real Material Fitting}
\begin{figure}[t]
\includegraphics[trim=2cm 1.2cm 2cm 0.8cm,clip, width=.9\linewidth]{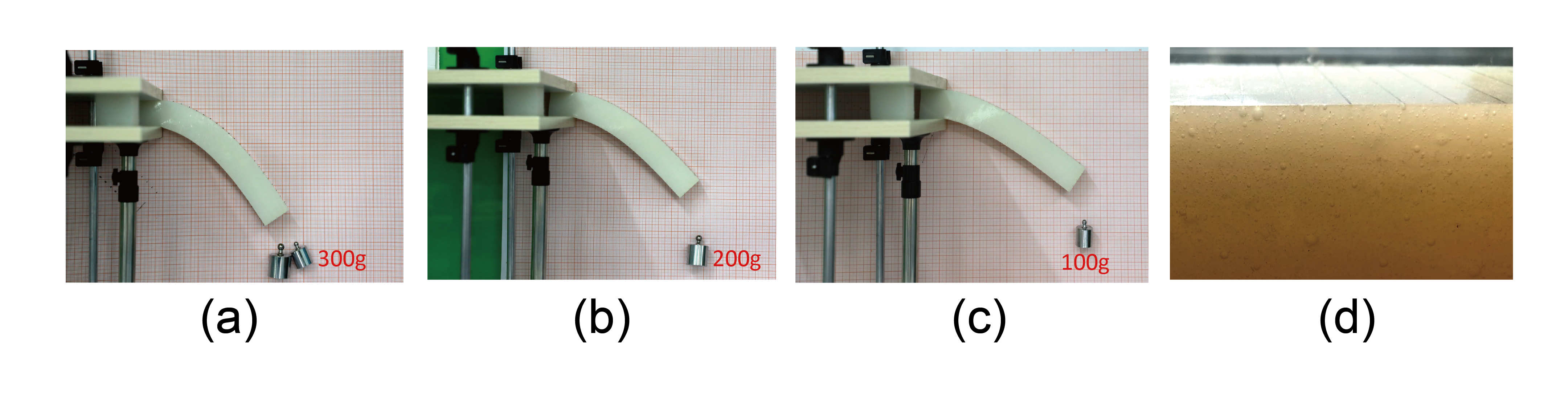}
\caption{Silicon bar examples: (a) $300$~g external loading case is used as training data, (b) $200$~g external loading case, and (c) $100$~g external loading case are used as test data; (d) irregular bubbles can be observed \pk{in a magnified view}. } 
\label{fig:Cantilever} 
\end{figure}

\binnew{We first validate our algorithm on a silicon bar example. The \pk{object} is made by casting \pk{an} elastomer material of type Silicon-601 into a bar-shaped mold. \pk{The} target material properties depend on the amount of added curing agent, }\bincheck{which \urin{has} no default nominal value.} \binnew{Moreover, \pk{the numerous small} air bubbles seen \pk{in} Fig.~\ref{fig:Cantilever}(d) add more variance \pk{to} the material properties. 
\bincheck{\pk{Thus, in the presence of these bubbles, it is} \urin{worthwhile} to 
\urin{allow} an algorithm like ours \urin{to} infer \pk{the} material properties.} As illustrated in Fig.~\ref{fig:Cantilever}(a) \pk{through} (c), the \pk{object} is fixed at one end, and three different external loads (100~g, 200~g, and 300~g) are added at the free end. We \pk{released} the load and \pk{recorded} its \pk{vibration}. The \pk{trajectory} of \pk{the} 300~g \pk{case} is used as training data, while the other two are used \pk{for testing}.} 
\pk{We learn the material using both our method and and the method of Wang et al.~\cite{Wang2015}.}
\bin{We encourage the reader to watch \pk{the} side by side comparison in the accompanying video.
}

\begin{figure}[!t]
\includegraphics[width=.9\linewidth]{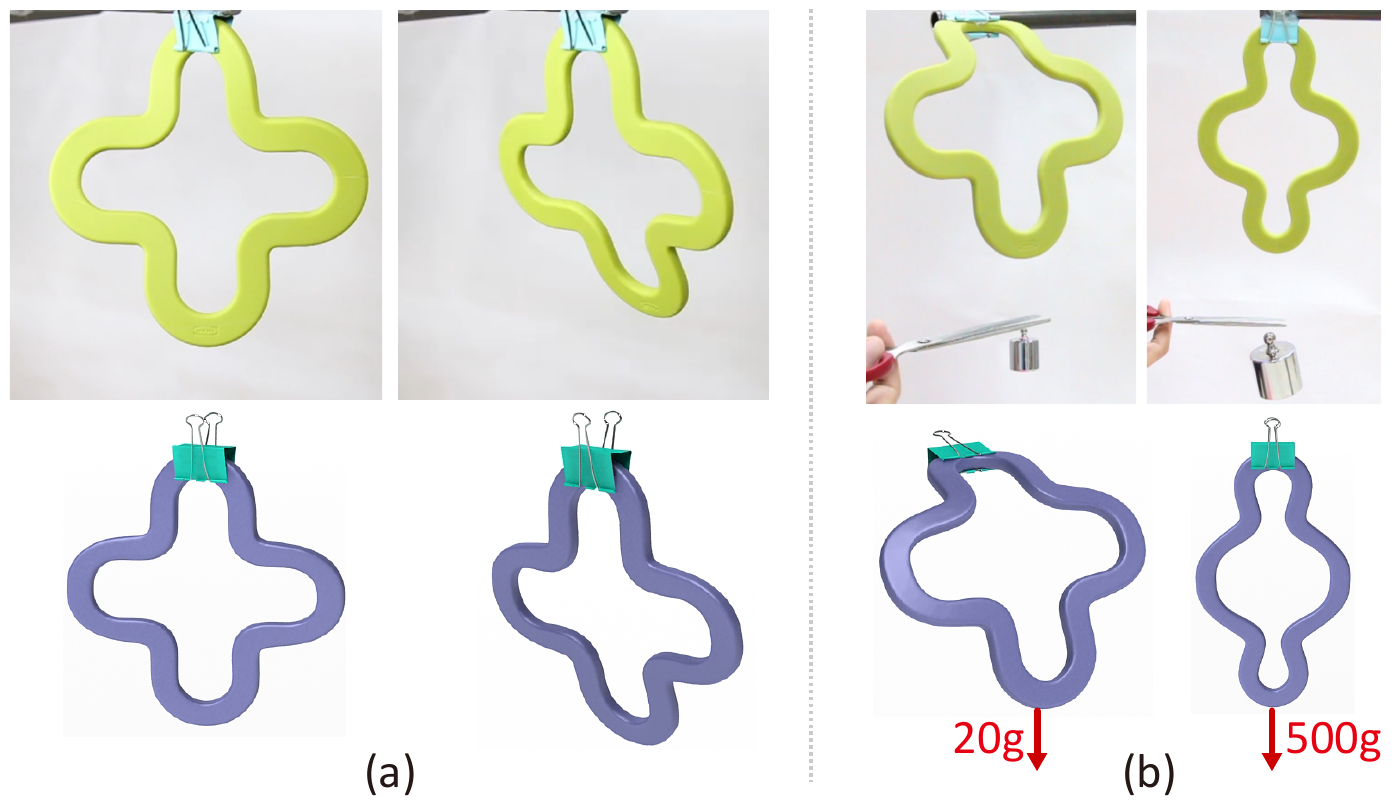}
\caption{Real Material Fitting: (a) Two captured trajectories and \urin{the corresponding} tracking result. (b) Static loading tests: a silicon pot holder is bent and pulled by external weights; the holder is fixed at one end horizontally and vertically. } 
\label{fig:realpotholder} 


\vspace{15pt}
\includegraphics[trim=0.5cm 1.2cm 2cm 0.8cm, width=.85\linewidth]{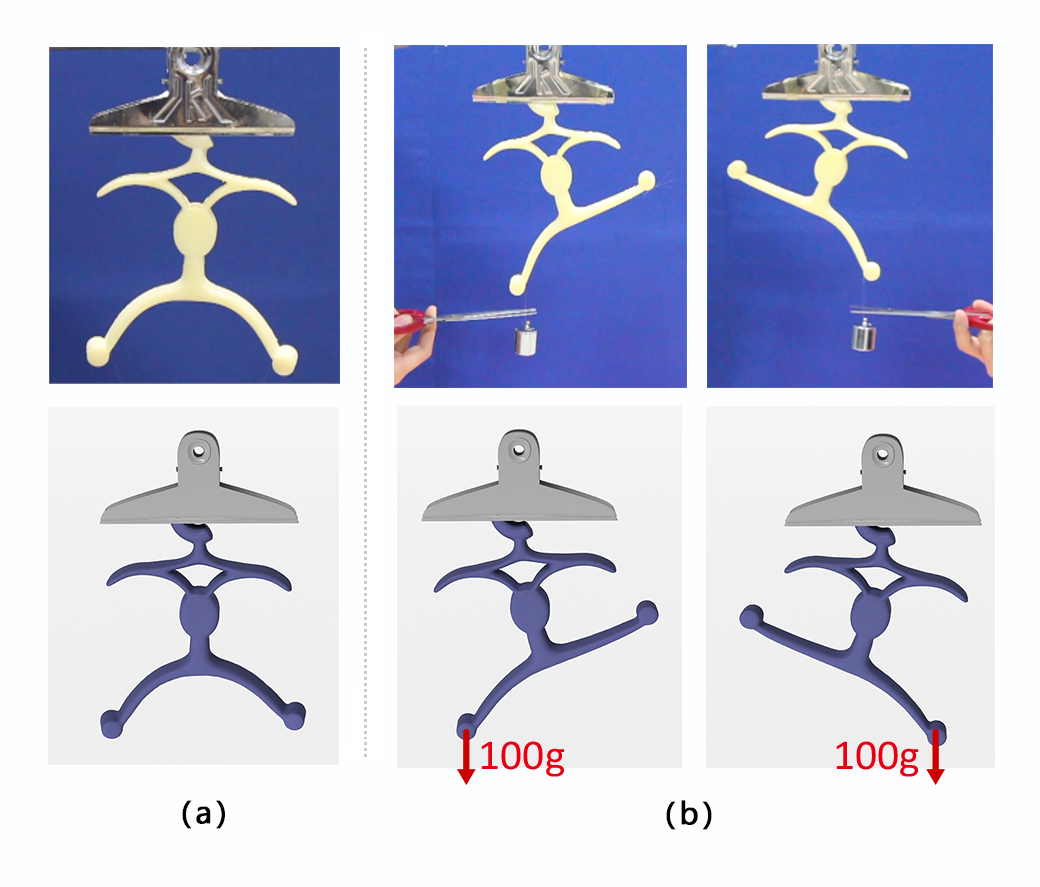}
\caption{Real Material Fitting: (a) One captured trajectories and \urin{the corresponding} tracking result. (b) Static loading tests: a silicon hanger is pulled by external weights; the hanger is fixed at top. } 
\label{fig:realhanger} 
\vspace{-5pt}
\end{figure}

\binnew{We also \pk{validated} and compared our algorithm with Wang et al.~\cite{Wang2015} \pk{for the} real silicon pot holder and hanger examples which are shown in \urin{Figs.~\ref{fig:realpotholder} and~\ref{fig:realhanger}, respectively}. \pk{The} captured trajectories \pk{seen in (a) of} both figures are used to obtain material corrections. }
\bin{The raw point cloud data are fused, and severe outliers are removed using Artec Studio.}
The results are validated through external loading tests. More specifically, we fix the objects at one end, either horizontally or vertically, and attach different weights at the other end. The external weights are suddenly released and the vibrations of the soft object are simulated and compared with the ground truth.
A side by side comparison can be seen in the accompanying video.
\binnew{For the hanger and silicon bar example, \pk{we} observed that the original constant mass damping coefficient \pk{$\alpha_0$}
\pk{must} be substituted with a polynomial function of principal stretch $\lambda$. }

\begin{table}[t]
\begin{center}
\caption{Performance statistics measured for different testing cases. From left to right: the test object, number of vertices, number of tet elements, number of frames for training data, number of reduced modes, number of learning iterations, number of RBF kernels,  and total computation time in hours for material learning. }
\label{tab:accuracy}
\small
\setlength{\tabcolsep}{3.5pt}
\begin{tabular}{|l|r|r|r|r|r|r|r|}
\hline
Case     & \#vert 	& \#tet & \#frm 	& \#mode 	& \#iter 	& \#kernel   & CPU  \\
\hline
Turtle   		 &347		&1185    & 400    	& 200     	& 29	& 46	& 0.9  	\\
Dragon   		 &959    	&2590    & 400    	& 200     	& 19    & 140   & 1.1  \\
Sphere 1   		 &2655    	&12712   & 400    	& 200     	& 27    & 140   & 9.1  \\
Sphere 2  		 &2655    	&12712   & 400    	& 200     	& 23    & 140   & 7.7 \\
Bar (Damp1)   &425    	&1536    & 300    	& 150     	& 7		& 400   & 0.4  \\
Bar (Damp2)   &425    	&1536    & 800    	& 400     	& 18   	& 500   & 2.5 \\
Pot Holder   	 &3031    	&8843    & 400    	& 200     	& 10    & 140   & 4.2 \\
\binnew{Hanger}   	 &1740    	&5888    & 375    	& 200     	& 4    & 500   & 2.0 \\
\binnew{Silicon Bar}   	 &650    	&2400    & 600    	& full     	& 6   & 300   & 0.5 \\
\hline
\end{tabular}
\end{center}
\vspace{-7pt}
\end{table}
\begin{figure}[t]
\includegraphics[width=.9\linewidth]{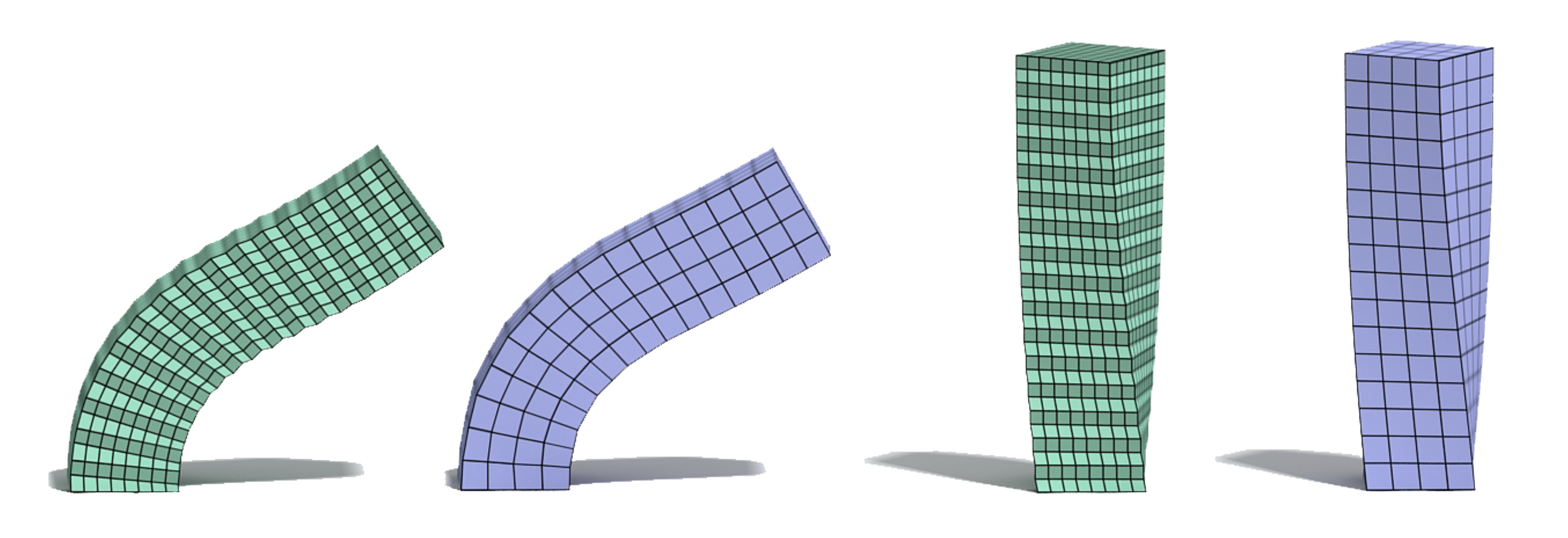}
\caption{Material coarsening. The green bar shows the fine mesh with a layered material distribution; 
the purple bar is the corresponding coarsened mesh with homogeneous material distribution. Bend (left) and twist (right) deformation trajectories of fine mesh are used as training data, and the purple bars are the reconstruction result\urin{s} after learning. }
\label{fig:Coarsening}
\end{figure}

\subsection{Material Coarsening}
The algorithm proposed in this paper can also be used for material coarsening.  In Fig.~\ref{fig:Coarsening}, a high resolution bar $(8\times8\times34)$ is composed of two different constitutive materials, with Young modulus values of 1e5 and 1e7, respectively. The two materials are composited in a layer by layer manner, represented by the light and dark green colors. The low resolution mesh is the result of coarsening by factor 2 along three axis directions. Two principal deformation modes (bend and twist) are used as training data. The equivalent coarsened material property found by our algorithm can produce very similar motion to the original high resolution heterogeneous model.


%

\subsection{Performance}
We \uri{measured} the \uri{computational} cost for each critical step on a 10-core 3.0 GHz Intel i7-6950X desktop. The performance for space-time optimization, listed in Table~\ref{tab:accuracy}, correlates with the number of tetrahedral elements, the number of frames in the motion trajectory, the number of RBF kernels, 
and the \urin{dimension} of selected reduced basis.

\section{Conclusion and Future Work}\label{sec:future}

We have presented a new method for estimating nonlinear constitutive models from trajectories of surface data.  
The key insight is to have a parametric material correction model learn the error of the elastic and damping properties of a nominal material.  
A framework for gradually learning this correction from only kinematic data is described. 
\bin{
We \urin{have} demonstrated our method with \urin{several} 
examples,  illustrating the ability of our approach to learn classical material models, user designed materials (cartoon physics), and real world captured data.
}

The desire to work with realistic constitutive models when simulating complex motion has been shared 
\urin{for a long while} by researchers from many fields, 
not just computer graphics. The possibility of employing machine learning towards such a goal is tantalizing. 
\bin{
We solved an interesting and timely problem which is critical for all machine learning algorithms 
\urin{for} generating annotated data automatically. 
We believe our present work is an important step in that direction.  
}


There are a number of interesting ideas to explore in future research.  
First, we note that extending our approach to accommodate a variety of numerical integration techniques would help 
\urin{in avoiding or reducing}
step size dependent numerical damping effects in our results. Second, we only address heterogeneous materials in the case of numerical coarsening to a homogeneous material.  
There are interesting extensions \urin{that can be considered}
for dealing with varying material properties across a model, for instance, by adding a latent material parameter to our representation.
\rev{Also, note that our framework is compatible with many other parametric representations.
For instance, a spline-based representation~\cite{Xu2015} could be another promising option to explore. }  
Finally, 
there are still a  variety of
potential damping effects that we cannot capture with \pk{our} approach.  The models we estimate do not account for any hysteresis in the damping model, while this can be common in real materials, as can also be the presence of plastic deformation.  Capturing a larger variety of complex plastic and  damping behaviors is indeed a very interesting avenue for future work.  

\section*{Acknowledgments}
We sincerely thank the reviewers for their constructive comments. This work was supported in part by National Key R\&D Program (2019YFF0302902, 2018YFB1403900), NSFC (61861130365, 61761146002), GD Talent Program (2019JC05X328), GD Science and Technology Program (2020A0505100064, 2018KZDXM058), and
NSERC Canada Discovery Grants (2018-05665, 84306).


\printbibliography                


\end{document}